\documentclass[fleqn,usenatbib,twocolumn]{mnras}

\usepackage[T1]{fontenc}
\usepackage{ae,aecompl}

\usepackage{graphicx,xcolor}	% Including figure files
\usepackage{amsmath}	% Advanced maths commands
\usepackage{amssymb}	% Extra maths symbols
\usepackage{threeparttable}

\usepackage{newtxtext,newtxmath}
\title[Chemical Enrichment in the Centaurus Cluster]{Chemical enrichment in the cool core of the Centaurus cluster of galaxies}

\author[K. Fukushima, S. B. Kobayashi, and K. Matsushita]{
Kotaro Fukushima$^{1}$\thanks{E-mail: kxfukushima@gmail.com, (KF)},
Shogo B. Kobayashi$^{1}$,
Kyoko Matsushita$^{1}$
\\
$^{1}$Department of Physics, Tokyo University of Science, 1-3 Kagurazaka,
Shinjuku-ku, Tokyo 162-8601, Japan\\
}

\date{Accepted XXX. Received YYY; in original form ZZZ}

\pubyear{2022}

\begin{document}
\label{firstpage}
\pagerange{\pageref{firstpage}--\pageref{lastpage}}
\maketitle

\begin{abstract}
Here we present results from over 500 kiloseconds
{\it Chandra} and {\it XMM-Newton} observations of the cool core of the Centaurus cluster.
We investigate the spatial distributions of the O, Mg, Si, S, Ar, Ca, Cr, Mn, Fe, and Ni abundances
in the intracluster medium with CCD detectors, and those of N, O, Ne, Mg,  Fe, and Ni
with the Reflection Grating Spectrometer (RGS).
The abundances of most of the elements show a sharp drop within the central 18\,arcsec,
although different detectors and atomic codes give significantly different values.
The abundance ratios of the above elements, including Ne/Fe with RGS,
show relatively flat radial distributions.
In the innermost regions with the dominant Fe-L lines,
the measurements of the absolute abundances are challenging.
For example, AtomDB and SPEXACT give Fe $=$ 0.5
and 1.4\,solar, respectively, for the spectra from the innermost region.
These results suggest some systematic uncertainties in the atomic data
and response matrices at least partly cause the abundance drop
rather than the metal depletion into the cold dust.
Except for super-solar N/Fe and Ni/Fe, sub-solar Ne/Fe, and Mg/Fe,
the abundance pattern agrees with the solar composition.
The entire pattern is challenging to reproduce
with the latest supernova nucleosynthesis models.
Observed super-solar N/O and comparable Mg abundance to stellar metallicity profiles
imply the mass-loss winds dominate the intracluster medium in the brightest cluster galaxy.
The solar Cr/Fe and Mn/Fe ratios indicate a significant contribution
of near- and sub-Chandrasekhar mass explosions of Type Ia supernovae.
\end{abstract}


%%%%%%%%%%%%%%%%%%%%%%%%%%%%%%%%%%%%%%%%%%%%%%%%%%

%%%%%%%%%%%%%%%%% BODY OF PAPER %%%%%%%%%%%%%%%%%%

\section{INTRODUCTION}
\label{sec:intro}

Elemental abundances in stars and gas in the Universe provide stringent constraints
on the formation and evolutionary history of galaxies.
Excluding primordial elements in the Universe like H, He, Li, and Be,
other heavy chemical elements were synthesised in stars and
expelled into interstellar space by supernovae (SNe).
Light $\alpha$-elements (O, Ne, Mg) mainly originate from
core-collapse SNe (CCSNe), and Fe-peak ones (Cr, Mn, Fe, Ni) are,
on the other hand, produced by thermonuclear explosions
in Type Ia SNe (SNeIa). The intermediate-mass elements (IMEs),
i.e. Si, S, Ar, and Ca, are synthesised by both SNeIa and CCSNe \citep[e.g.,][and references therein]{Nomoto13}.
Unlike these elements,  N is mainly synthesised
in low- or intermediate-mass stars and expelled into the interstellar medium (ISM)
through stellar mass loss \citep[e.g.,][]{Nomoto13}.

The intracluster medium (ICM), hot X-ray emitting plasma
pervading the galaxy clusters,
retains a dominant fraction of heavy elements
(e.g., N, O, Ne, Mg, Si, S, Ar, Ca, Cr, Mn, Fe, and Ni)
synthesised by stars and SN explosions in member galaxies.
The abundance of these elements can be well constrained
from the intensity of their K-shell emission lines
within the X-ray band \citep[e.g.,][for a recent review]{Mernier18c}.
Therefore, X-ray spectroscopy of the ICM is one of
the most reliable ways to investigate the chemical
enrichment history in clusters \citep[e.g.,][]{dePlaa07,Sato07,Mernier18a}.
In the last few decades, spatially resolved spectra of
the ICM observed with modern X-ray observatories like
{\it Chandra}, {\it XMM-Newton} and {\it Suzaku} have allowed us to study 
the spatial distribution of metals in the ICM.
Outside the core regions, flat and azimuthally uniform Fe distributions towards the outskirts
have been reported \citep[e.g.,][]{Matsushita11,Matsushita13a,Werner13,Simionescu15,Urban17}.
This remarkably extended Fe distribution requires the early enrichment of the ICM
before the cluster formation, i.e. ten billion years ago,
also supported by cosmological simulations \citep[e.g.,][]{Biffi18}.

X-ray luminous clusters showing a strong
surface brightness peak towards the centre,
i.e. cool-core clusters,  are ideal targets to study 
ongoing enrichment process by the brightest cluster
galaxies (BCGs). Most of them are classified as elliptical, and 
located in the X-ray peak of the cool cores.
Observational studies of the metal abundances in cool cores
provide a powerful probe for where and when metals were dispersed into the ICM.
A central Fe abundance excess within the ICM was first reported
in the Centaurus cluster by {\it ROSAT} and {\it ASCA} \citep{AF94,Fukazawa94}.
Many other cool-core clusters exhibit similar trends for metal abundances
\citep[e.g.,][]{deGrandi01,Million11,Mernier17}.
To produce the central Fe abundance peaks, \citet{Boehringer04} proposed 
enrichment by SNeIa over a relatively long time ($> 5$\,Gyr).
In addition, metal enrichment by CCSNe when the cluster galaxies were still actively star-forming
has been  discussed based on central peaks of $\alpha$-elements
\citep[e.g.,][]{dePlaa06,Simionescu09a,Mernier17,Erdim21}.
Recent observation in the Perseus cluster core
by {\it Hitomi} provided compelling results that
the O/Fe, Ne/Fe, Mg/Fe, Si/Fe, S/Fe, Ar/Fe, Ca/Fe,
Cr/Fe, Mn/Fe, and Ni/Fe abundance ratios in the ICM
are fully consistent with those in the solar,
and therefore the Milky Way \citep{Hitomi17,Simionescu19}.
This solar chemical composition in the Perseus core
requires not only CCSN but also near- and sub-Chandrasekhar mass ($M_\textrm{Ch}$)
SNIa contribution to the enrichment of the ICM.
Although the abundance ratio pattern could not be
reproduced sufficiently by the latest SN nucleosynthesis models,
the study of the Perseus core provided an important clue to
chemical enrichment in the cluster centre and possibly BCG.

Within a central few-kiloparsec,
the situation is more puzzling and fascinating in the innermost core.
Since the first discovery in A2199 \citep{Johnstone02},
central abundance drops are reported especially for Fe
in some cool-core clusters wherein the enhanced enrichment
is expected from the central BCG \citep[e.g.,][]{Churazov03,Panagoulia15,Liu19b}. 
The central abundance drop in the Centaurus cluster was first discovered
by \citet{SF02} with a remarkable depletion
from $\sim$\,2 to $\sim$\,0.5\,solar towards the centre.
With optical and infrared observations, cold dust filaments are detected
in BCGs \citep[e.g.,][for the Centaurus cluster]{Crawford05,Mittal11}, and 
\citet{Panagoulia15} proposed that these abundance drops would be caused by 
depletion of metals into the cool dust grains in the BCG.
\citet{Lakhchaura19} reported that the abundance of the `non-reactive' element
Ar shows a relatively slight central drop,
while Si and S show a remarkable abundance drop as Fe.
The abundance drops may originate from the feedback of active galactic nuclei
which would dissipate a fraction of central metal-rich gas
towards the outer radius \citep[e.g.,][]{Sanders16,Liu19b}.

The Centaurus cluster, also known as A3526,
is a well-known cool-core cluster,
within which NGC~4696 is residing as the BCG.
Because being a nearby and X-ray luminous galaxy cluster,
the Centaurus cluster is one of the most compelling examples as well as
the Perseus cluster for studies of the metal abundance in the ICM.
It has been reported that the abundance in a few-arcmin core
of the Centaurus cluster shows super-solar ($\sim$\,1.5--2\,solar) value for IMEs and Fe
by X-ray missions like {\it Chandra}, {\it XMM-Newton}, and {\it Suzaku}
\citep[e.g.,][]{Matsushita07,Takahashi09,Sakuma11,Sanders16}.
The Fe abundance in other cluster cores, on the other hand,
is typically sub-solar ($\sim$\,0.8\,solar), which is close to  those 
of groups and early-type galaxies \citep[e.g.,][]{Konami14,Mernier18b}.
Then, the Centaurus cluster core do be the specific but ideal target
for a comprehensive study of SNIa contribution to the enrichment
due to the high Fe abundance.

In this paper, we analyse the deepest CCD data with {\it Chandra} and {\it XMM-Newton}
to date of the cool core of the Centaurus cluster and NGC~4696.
We study the spatial distributions of metal abundances of O, Mg, IMEs, and 
Fe-peak elements, including Cr, Mn, and Ni, to constrain the metal enrichment history by the BCG.
In addition, we analyse grating data onboard {\it XMM-Newton}
to study the spatial distribution of Ne and N.
Due to its non-chemical reactivity, radial distribution of the Ne abundance
would be crucial to restrict the mechanism of abundance drops at the cluster centre.
We also make a comparison between the latest versions of
two different atomic codes, the atomic data base (AtomDB, \citealt[][]{Foster12})
and the {\sc spex} Atomic Code and Tables (SPEXACT, \citealt[][]{Kaastra96}),
which have been updated in response to {\it Hitomi} \citep[][]{Hitomi18b}.
This paper is organized as follows. In Section\,\ref{sec:observation},
we describe details of our {\it Chandra} and {\it XMM-Newton} observations and data reduction.
In Section\,\ref{sec:results}, we present our spectral analysis.
We discuss the results in Section\,\ref{sec:discussion} and
present our conclusions in Section\,\ref{sec:conclusions}.
In this paper, we assume cosmological parameters as
$H_0=70$\,km\,s$^{-1}$\,Mpc$^{-1}$, $\Omega_m=0.3$ and $\Omega_{\Lambda}=0.7$,
for which 1\,arcsec corresponds to $\sim$\,0.2\,kpc at the redshift
of 0.0114 for NGC~4696 \citep[][]{SR99}.
All abundances in this work are relative to the proto-solar values of \citet{Lodders09}.
The errors are at 1$\sigma$ confidence level unless otherwise stated.

\section{OBSERVATIONS AND DATA REDUCTION}
\label{sec:observation}

We used the publicly archival observation data of the core
region of the Centaurus cluster with the {\it Chandra}
Advanced CCD Imaging Spectrometer (ACIS, \citealt{Weisskopf02}),
the European Photon Imaging Camera (EPIC, \citealt{Turner01,Strueder01}),
and Reflecting Grating Spectrometer (RGS, \citealt{denHerder01}) on-board {\it XMM-Newton}.
Table\,\ref{tab:observation} lists the observations used in this work.

\begin{table}
\centering
\caption{List of observations used in this work \label{tab:observation}}

\begin{threeparttable}
\begin{tabular}{@{}c@{\:\:\:}c@{\:\:\:}c@{\:\:\:}c@{\:\:\:}c}\\ \hline\hline
Obs. ID & Instrument & Cleaned Exposure & Offset\tnote{a} & Date \\
 & & (ks) & (arcmin) & \\ \hline
\multicolumn{5}{c}{{\it Chandra}} \\ \hline
504 & ACIS-S & 20.85 & 0.13 & 2000 May 22 \\
505 & ACIS-S & 8.97 & 0.13 & 2000 Jun 08 \\
4190 & ACIS-S & 30.52 & 3.79 & 2003 Apr 18 \\
4191 & ACIS-S & 30.20 & 4.23 & 2003 Apr 18 \\
4954 & ACIS-S & 73.73 & 0.10 & 2004 Apr 01 \\
4955 & ACIS-S & 36.33 & 0.10 & 2004 Apr 02 \\
5310 & ACIS-S & 40.60 & 0.10 & 2004 Apr 04 \\
16223 & ACIS-S & 144.72 & 0.10 & 2014 May 26 \\
16224 & ACIS-S & 29.40 & 0.10 & 2014 Apr 09 \\
16225 & ACIS-S & 28.32 & 0.10 & 2014 Apr 26 \\
16534 & ACIS-S & 47.17 & 0.10 & 2014 Jun 05 \\
16607 & ACIS-S & 39.92 & 0.10 & 2014 Apr 12 \\
16608 & ACIS-S & 23.09 & 0.10 & 2014 Apr 07 \\
16609 & ACIS-S & 71.97 & 0.10 & 2014 May 04 \\
16610 & ACIS-S & 16.43 & 0.10 & 2014 Apr 27 \\ \hline
\multicolumn{5}{c}{{\it XMM-Newton}} \\ \hline
0046340101 & MOS1 & 43.26 & 0.002 & 2002 Jan 03 \\
 & MOS2 & 43.55 & \\
 & pn & 30.46 & \\
 & RGS1 & 44.83 & \\
 & RGS2 & 43.51 & \\
0406200101 & MOS1 & 100.91 & 0.11 & 2007 Aug 22 \\
 & MOS2 & 103.69 & \\
 & pn & 71.88 & \\
 & RGS1 & 101.58 & \\
 & RGS2 & 101.58 & \\
0823580101 & MOS1 & 102.26 & 10.7 & 2018 Jul 03 \\
  & MOS2 & 104.39 & \\
  & pn & 104.39 & \\
0823580201 & MOS1 & 100.77 & 11.6 & 2018 Aug 09 \\
  & MOS2 & 102.79 & \\
  & pn & 76.06 & \\
0823580501 & MOS1 & 96.85 & 10.3 & 2018 Jul 20 \\
  & MOS2 & 98.37 & \\
  & pn & 72.65 & \\
0823580601 & MOS1 & 99.88 & 11.3 & 2018 Jul 30 \\
  & MOS2 & 99.78 & \\
  & pn & 74.25 & \\ \hline
\end{tabular}

\begin{tablenotes}
\item[a] The angular distance from NGC~4696 to the target coordinates of each observation.
\end{tablenotes}

\end{threeparttable}
\end{table}

\begin{figure*}
\includegraphics[width=\textwidth]{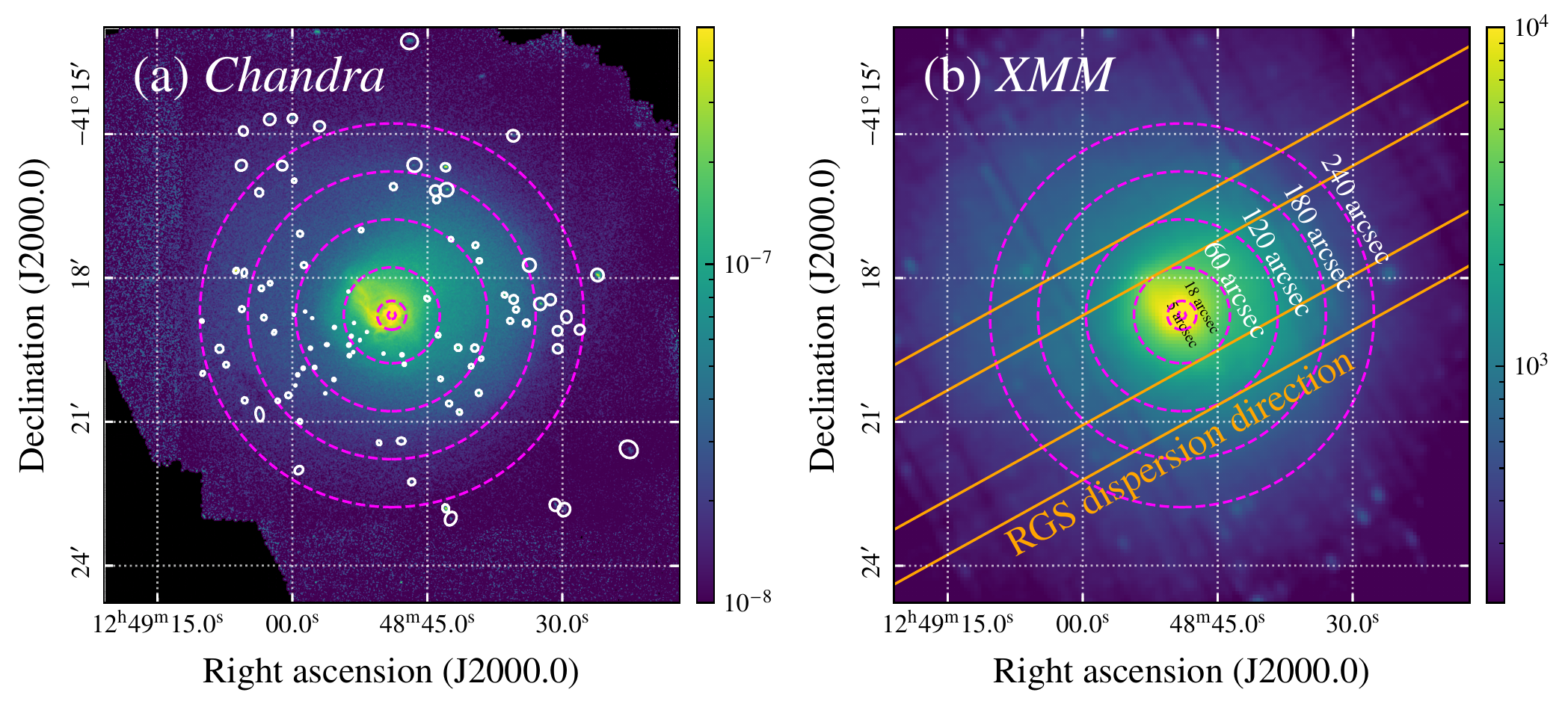}
\caption{Images of the Centaurus cluster core in the $0.5$--$7.0$\,keV band.
(a) Vignetting corrected image with the {\it Chandra} ACIS-S,
where the colour bar indicates photon flux in cm$^{-2}$\,s$^{-1}$.
The magenta dashed annuli represent 5, 18, 60, 120, 180,
and 240\,arcsec spectrum extraction regions. The white ellipses
show the detected point sources (see \S\,\ref{subsec:chandra_reduction} for details).
(b) A mosaiced count image with the {\it XMM-Newton} EPIC MOS and pn.
The extraction region for RGS spectra (60--120\,arcsec width on both sides, for instance)
is shown with the orange lines, where the cross-dispersion width
of RGS is between 60 and 120\,arcsec lines.
\label{fig:image}}
\end{figure*}

\subsection{{\it Chandra}}
\label{subsec:chandra_reduction}

We used {\it Chandra} Interactive Analysis 
of Observations ({\sc ciao}) version 4.12 \citep{Fruscione06}
and calibration database ({\sc caldb}) version 4.8.4.1 for
the following data reduction.
We reprocessed the data with {\tt chandra\_repro} tool keeping the standard procedure by {\it Chandra} team, 
and remove strong background flares using {\tt deflare} script in the {\sc ciao} package.
The cleaned ACIS data have a total exposure of 640\,ks after the procedures stated above.
We also searched for faint point sources using the {\tt wavdetect} algorithm
in the energy range of 0.5--7.0\,keV (encircled counts fraction $=0.9$, threshold significance $=10^{-6}$),
excluding sources detected at the cluster centre and
the surrounding ICM (e.g., plume-like structure, \citealt{Sanders16}).
We extracted ACIS spectra from the annular regions shown in Fig.\,\ref{fig:image}(a)
and generated redistribution matrix files (RMFs) and ancillary response files (ARFs) for each ObsID.
The outermost radius of our extracting regions (240\,arcsec) is about
three times the effective radius ($R_\textrm{e}$) of NGC~4696 ($\sim$\,85\,arcsec, \citealt{Carollo93}),
indicating that most of stars in NGC~4696 are within our extraction regions.
In addition, this aperture is close to a radius at which the ICM temperature
starts to decline towards the centre \citep[][]{Sanders16}.
The spectra and response files taken in the same observation years are coadded to improve the photon statistics.
Background spectra are extracted from all field of view of ACIS-S3 chip
out of 3\,arcmin core centred on the cluster centre.

\subsection{{\it XMM-Newton}}
\label{subsec:xmm_reduction}

\subsubsection{EPIC}
\label{subsubsec:xmm_ccd}

The reduction procedure for the data sets of {\it XMM-Newton} was
performed using the {\it XMM-Newton} Science Analysis System 
({\sc sas}) version 18.0.0. We reprocessed and filtered all the EPIC data
using {\tt emchain} and {\tt epchain} for the MOS and pn, respectively.
The excluding sources are taken from the {\it Chandra} ACIS detection
(Fig.\,\ref{fig:image}(a)) at the same position and radius.
All reprocess procedures are in accordance with the standard reduction criteria by {\sc sas}.
After the reprocess, the total exposure times are 550\,ks (MOS), and 380\,ks (pn).
For each observation, we extracted EPIC spectra from the same
annular regions as {\it Chandra} (Fig.\,\ref{fig:image}(b)) and generated
RMFs and ARFs for each ObsID,
but did not merge them because the aim points
of each observation are significantly shifted.
We fit the spectra of MOS1 and MOS2 jointly with the same spectral parameters.
The background spectra were extracted from the  annular region
with 7.5--11.7\,arcmin.

\subsubsection{RGS}
\label{subsubsec:xmm_rgs}

The RGS spectra provide us spatial information of extended emission
over 5\,arcmin width along the cross-dispersion direction
when the X-ray peak of the target locates near the pointing position \citep[e.g.,][]{Chen18,Zhang19}.
Thus, we used only two datasets in 2002 and 2007
which have quite smaller offsets than the others (see Table\,\ref{tab:observation}).
The RGS data were processed with {\tt rgsproc}.
After filtering flared events as well as the EPIC data,
the remaining total exposure time is 140\,ks.
As done by \citet{Chen18}, we extracted first- and second-order RGS spectra 
centred on the emission peak along the dispersion direction instead of
using the {\tt xpsfincl} which is generally used.
For each annular region within 120\,arcsec for the CCD spectra,
we filtered events by the cross-dispersion direction and extracted
spectra over two wide regions on both sides.
For instance, the two wide regions for 60--120\,arcsec are plotted in Fig.\,\ref{fig:image}(b).
The first- and second-order spectra are jointly fitted,
where the RGS1 and RGS2 spectra of each order are combined.

\section{ANALYSIS \& RESULTS}
\label{sec:results}

\subsection{Spectral Fitting}
\label{subsec:fitting}

\subsubsection{CCD spectra}
\label{subsubsec:ccd_fit}

\begin{figure*}
\includegraphics[width=\textwidth]{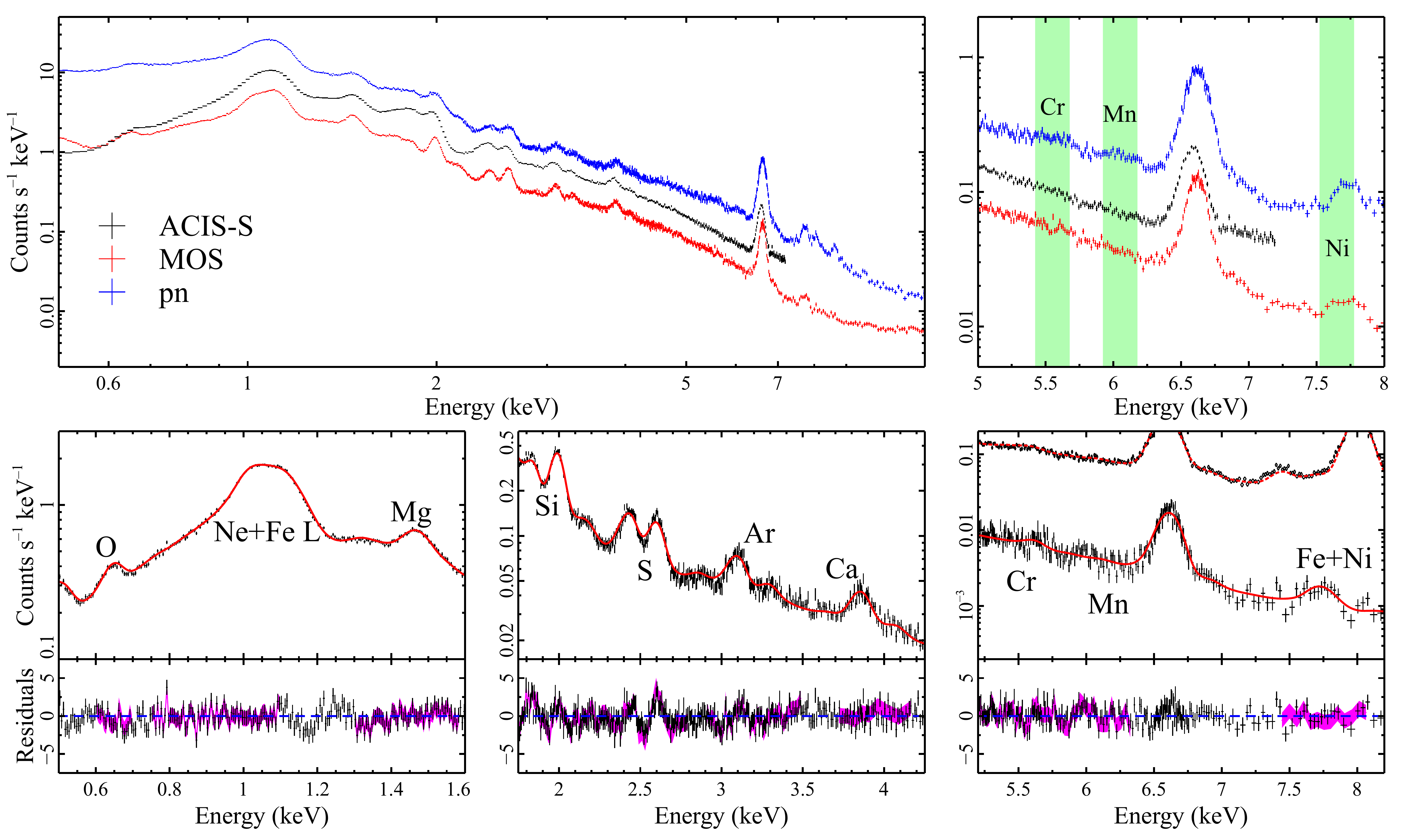}
\caption{Upper: The stacked spectra of the central 240\,arcsec
region for each CCD camera, all of which are folded
through the spectral response of each detector.
The black, red, and blue data represent ACIS-S, MOS, and pn spectra, respectively.
We also plot narrow band spectra in order to show the emission lines of Fe-peak elements.
Lower: The representative spectra extracted from 18--60\,arcsec annular region
for MOS and best-fitting models enlarging on prominent K-shell emission lines.
The black and magenta residuals represent
the global and local fit results, respectively
(see \S\,\ref{subsubsec:ccd_fit} for detail).
The pn spectrum is also plotted with diamonds for the Fe-peak elements.
Emission lines for some elements are marked.
\label{fig:ccd_spec}}
\end{figure*}

Here we analyse X-ray spectra extracted from
the annular regions given in Fig.\,\ref{fig:image} using {\sc xspec} package version 12.10.1f \citep{Arnaud96}.
We use the latest AtomDB \footnote{\url{http://www.atomdb.org}}
version 3.0.9 to model a collisional ionization equilibrium (CIE) plasma.
We model and fit the background spectra simultaneously
with the source spectra, instead of subtracting them from the source data.
The detector background model consists of a broken power-law
with Gaussian lines for the instrumental fluorescence
lines like Al-K, Si-K, and Cu-K \citep[e.g.,][]{LM08,Mernier15}.
We also take the background celestial emissions into account
with a power-law (1.4 photon index) for the cosmic X-ray background and
two CIE plasma components with solar metallicity for the local hot bubble (LHB)
and the Galactic thermal emission \citep[e.g.][]{Yoshino09}.
These components except for the LHB are modified by the Galactic absorption and
the temperature of the LHB is fixed at 0.1 keV.
The derived temperature of the Galactic thermal emission is $0.94\pm 0.02$\,keV,
indicating that the Centaurus cluster is located at a high-temperature region of
the Galactic emission \citep{Yoshino09, Nakashima18}.
Because the ICM of the Centaurus cluster is extended over several degrees,
the emission component from the ICM is also included in the spectral fitting model
for the background region. 
In the following spectral analysis, we fit all spectra using the C-statistics \citep[][]{Cash79},
to estimate the model parameters and their error ranges
without bias \citep{Kaastra17}.
The spectra are rebinned to have a minimum of 1 count per spectral bin.
The stacked spectra of the central 240\,arcsec region are shown
in the upper panel of Fig.\,\ref{fig:ccd_spec}.
Emission lines of O, Fe-L (+\,Ne), Mg, Si, S, Ar, Ca, Fe and Ni (+\,Fe)  are clearly seen in the spectra.
In addition, there are some hints of Cr and Mn lines in the pn spectrum.

\begin{figure}
\includegraphics[width=\columnwidth]{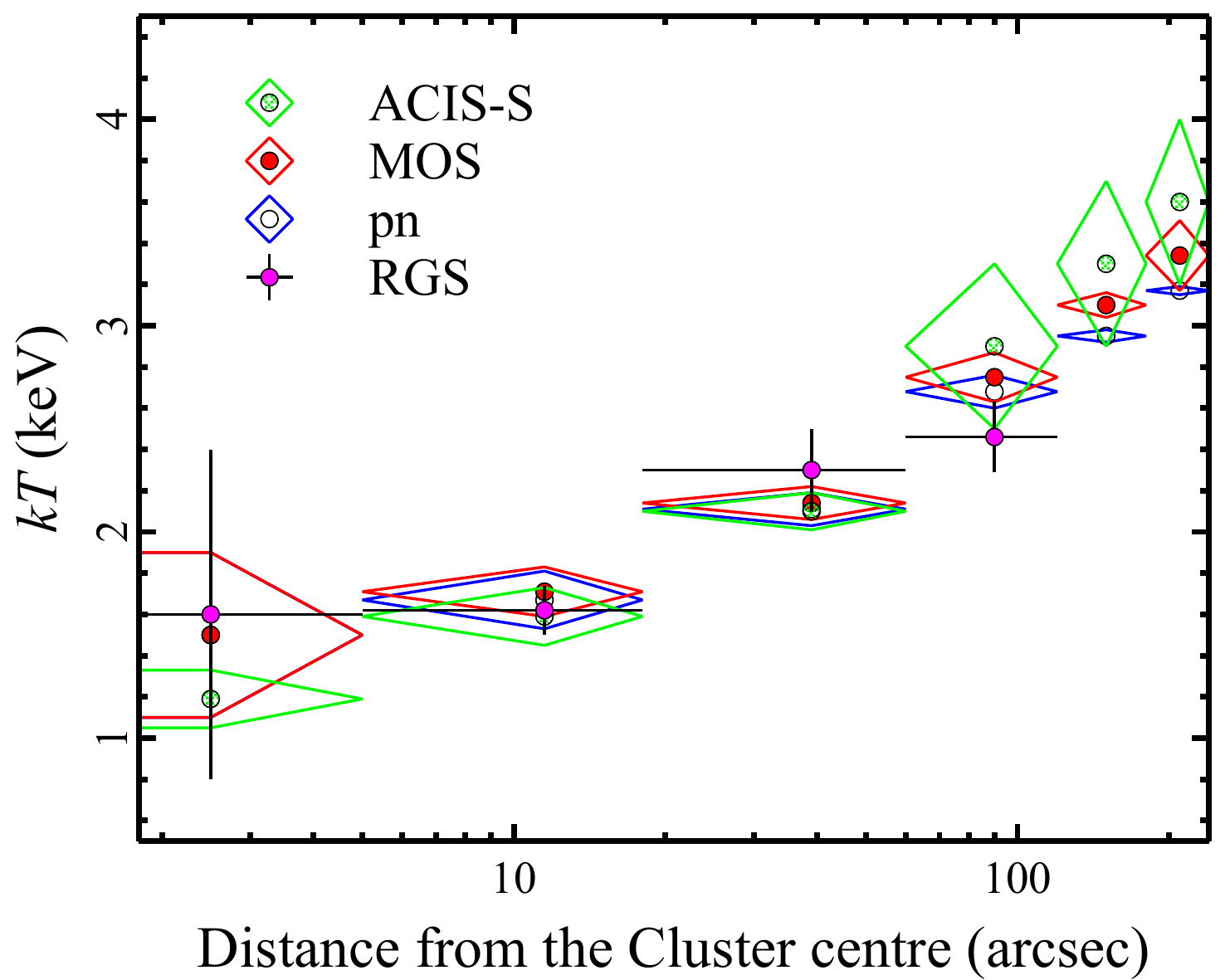}
\caption{Radial profile of the VEM-weighted average temperature.
The green-, red-, and blue-diamond plots are for results with {\it Chandra} ACIS-S, {\it XMM-Newton} MOS,
pn, respectively. The magenta plots indicate the RGS results.
\label{fig:kT_prof}}
\end{figure}

Systematic uncertainties in the response matrices and atomic codes
bias measurements of line and continuum temperatures.
Therefore, we adopt {\tt bvvtapec} model, which allows different temperatures
for the line and continuum \citep[e.g.,][for detailed discussion]{Hitomi18a}.
The ICM spectra are modelled with a triple {\tt bvvtapec} components
to consider the multi-temperature structure in the cool core.
We assume the three thermal components have the same metal abundances.
The Galactic absorption is modelled using the {\tt phabs} 
assuming the photoelectric absorption cross-sections taken from \citet{Verner96}.
We let the absorption column density $N_{\textrm{H}}$
and elemental abundance of O, Ne, Mg, Si, S, Ar, Ca,
Cr, Mn, Fe, and Ni vary freely.
The abundances of elements lighter than O are fixed to solar.
The other element abundances are tied to those of the nearest lower atomic number element,
which is allowed to vary.
For example, the F abundance is linked to that of O.
The line and continuum temperatures for the middle-
and coolest-temperature components cannot be well constrained independently
and thus are assumed to be halved and quartered the hottest-temperature ones, respectively.
We also let the volume emission measure (VEM) free to vary,
given as $\int n_e n_\textrm{H} dV/(4\pi D^2)$,
where $n_e$ and $n_\textrm{H}$ is electron and proton density,
$V$ and $D$ are the volume of the emission region
and the distance to the emitting source, respectively.
This model gives good fits to the spectra of the individual annular regions, yielding C-stat/dof
$\sim 1$--$1.3$ as summarised in Table\,\ref{tab:parameter_1} and Fig.\,\ref{fig:ccd_spec}. 
Since the differences between line and continuum
temperatures are small ($< 0.3$\,keV),
we hereafter report the continuum temperature as representative.
As shown in Fig.\,\ref{fig:kT_prof}, the different CCD detectors
give consistent VEM-weighted average temperatures (denoted $kT$) for each annular region.
This profile shows a peak temperature of $\sim$\,3\,keV
and a clear drop towards the centre, which is consistent with a more extended profile by \cite{Sanders16}.
Therefore, our extracting regions cover an important part of the cool core of the Centaurus cluster
as noted in \S\,\ref{subsec:chandra_reduction}.

\begin{figure*}
\includegraphics[width=\textwidth]{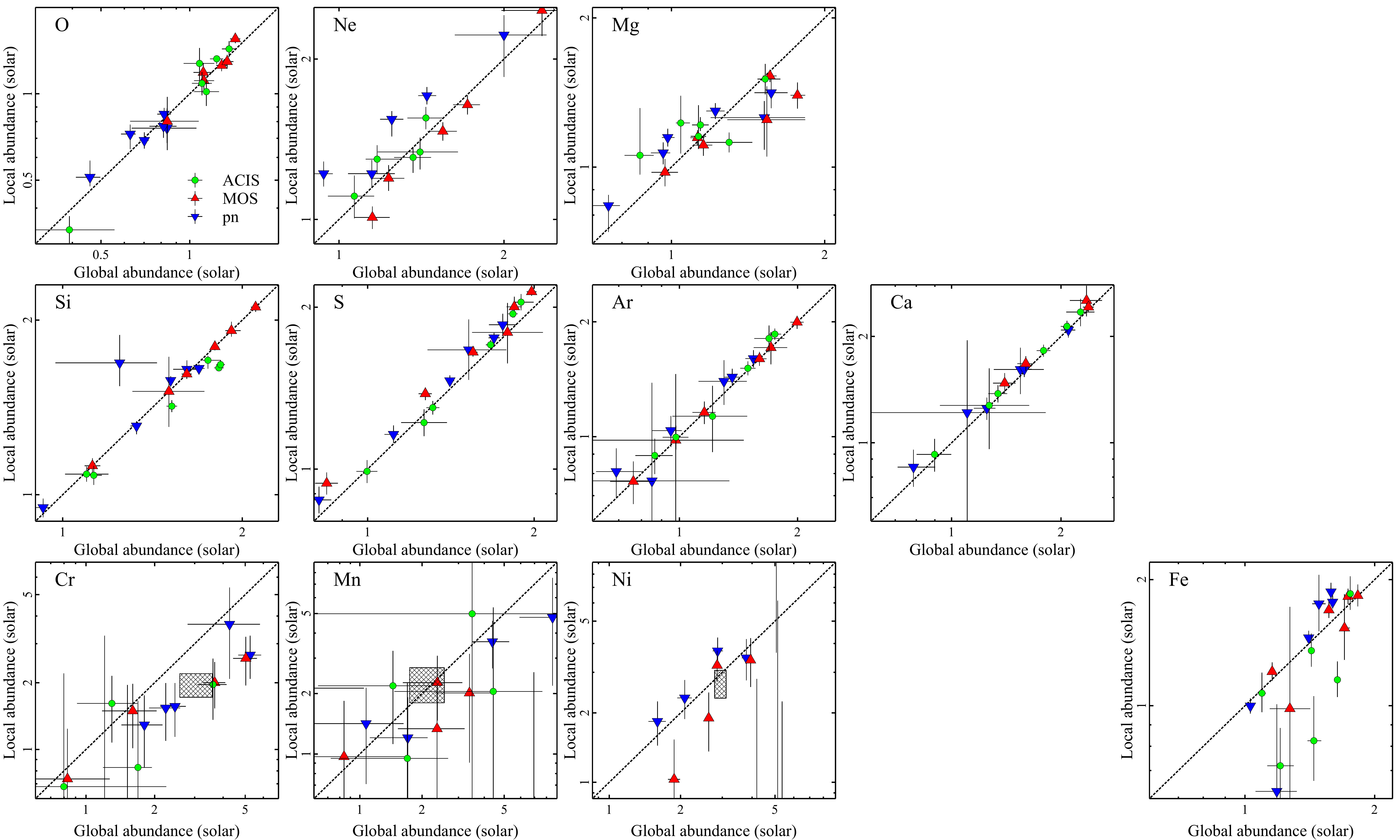}
\caption{Comparisons of the element abundances in each annular
region obtained from global fit with local one.
The circle plots indicate the values with ACIS-S3,
while the up- and down-triangle ones are for MOS and pn, respectively.
The shaded squares for Fe-peak elements are the mean
abundances of all CCDs measured in 18--120\,arcsec region.
\label{fig:abund_glo_vs_loc}}
\end{figure*}

Next, we re-fit the spectra within local energy bands
around the K-shell emission lines of the individual elements
in order to minimize biases due to the response uncertainty
on the metal abundance measurements,
i.e. we perform local or narrow-band fits
\citep[e.g.,][]{Lakhchaura19,Simionescu19}.
The ratios of each VEM are fixed to the value of global fit,
leaving the abundance of a considered element to vary freely and tying all
other parameters to the global fit values in Table\,\ref{tab:parameter_1}.
The boundaries of these local bands are 0.51--0.77\,keV for O,
0.75--1.25\,keV for Ne, 1.36--1.59\,keV for Mg, 
1.65--2.26\,keV for Si, 2.37--4.44\,keV for other intermediate-mass elements (IMEs), i.e. S, Ar, and Ca,
5.01--6.59\,keV for Cr and Mn, 6.25--7.25\,keV for Fe, and 7.24--8.02\,keV for Ni.
Here, to avoid contaminations from the strong instrumental Cu line,
we use the two observations pointing towards the cluster centre
for the local-fits around the Ni lines of the pn spectra.
The representative spectrum and the best-fitting residuals around each prominent line
are plotted in the lower panels of Fig.\,\ref{fig:ccd_spec}.
The obtained elemental abundances with the local fits are shown in
Table\,\ref{tab:parameter_2} and \ref{tab:parameter_3}.
In Fig.\,\ref{fig:abund_glo_vs_loc}, we compare the derived abundances
from the global and local fits. The abundance values and the residual structure
around each K-shell emission line with two fit methods are
globally similar except for the Cr abundances.
The Ne abundances with two fit methods are globally consistent,
except for the results with pn which offers a relatively lower spectral resolution than that of MOS.
However, the Ne~K-shell emission lines are blended in the Fe-L bump,
and thus, our estimation of the Ne abundances from the CCD spectra would be suffered
from some systematic uncertainties.
The Mg-K shell lines are also blended in the Fe-L bump.
The Mg abundances with pn and ACIS show a relatively large scatter,
while those with MOS, possibly the most reliable instrument among the three,
are consistent except for only one point.
We suspect that these discrepancies are mainly caused by
the moderate spectral resolution and/or sensitivity of the pn and ACIS instruments.
The local fits around the Fe-K lines mostly give consistent results with those from the global fits.
The exceptions are the three regions within 60\, arcsec with {\it Chandra} and the innermost region with pn,
where the local fits give even lower Fe abundances than the global fits.
Hereafter, we use the results of the global fits for the Fe abundance,
and those of the local fits for the other elements.

\begin{table*}
\centering
\caption{Best-fitting spectral parameters for the ICM emission from the global fits. \label{tab:parameter_1}}

\begin{threeparttable}
\begin{tabular}{ccccccccc} % manual @ spacing to prevent this being too wide for a page
\hline\hline
Radial range & $N_\textrm{H}$\tnote{a} & $kT_\textrm{hot}$ & $kT$\tnote{b} & VEM$_\textrm{hot}$\tnote{c} & VEM$_\textrm{middle}$ & VEM$_\textrm{cool}$  & Fe & C-stat/dof \\
(arcsec) & (10$^{20}$\,cm$^{-2}$) & (keV) & (keV) & (10$^{10}$\,cm$^{-5}$) & (10$^{10}$\,cm$^{-5}$) & (10$^{10}$\,cm$^{-5}$) & (solar) & \\
\hline
\multicolumn{8}{c}{{\it Chandra} ACIS-S}\\ \hline
$0$--$5$ & $9.65$\tnote{d} & $2.51\pm 0.08$ & $1.19\pm 0.14$ & $0.5\pm 0.1$ & $1.85\pm 0.06$ & $1.29\pm 0.03$ & $1.21\pm 0.09$ & 3233.5/2935\\
$5$--$18$ & $11.9\pm 0.6$ & $3.02\pm 0.09$ & $1.59\pm 0.14$ & $2.7\pm 0.3$ & $10.0\pm 0.5$ & $3.7\pm 0.1$ & $1.44\pm 0.06$ & 3690.4/3435\\
$18$--$60$ & $14.0\pm 0.2$ & $3.05\pm 0.04$ & $2.10\pm 0.09$ & $31\pm 1$ & $60\pm 1$ & $0.88\pm 0.07$ & $1.64\pm 0.02$ & 4476.1/3624 \\
$60$--$120$ & $13.2\pm 0.2$ & $4.81\pm 0.12$ & $2.9\pm 0.4$ & $25\pm 5$ & $94\pm 4$ & $<0.02$ & $1.76\pm 0.02$ & 4315.0/3636 \\
$120$--$180$ & $13.3\pm 0.2$ & $5.47\pm 0.22$ & $3.3\pm 0.4$ & $24\pm 4$ & $87\pm 4$ & $<0.04$ & $1.43\pm 0.02$ & 4246.0/3637 \\
$180$--$240$ & $14.8\pm 0.3$ & $5.43\pm 0.09$ & $3.6\pm 0.6$ & $30\pm 6$ & $59\pm 6$ & $<0.35$ & $1.09\pm 0.03$ & 4120.9/3637 \\ \hline
\multicolumn{8}{c}{{\it XMM-Newton} MOS}\\ \hline
$0$--$5$ & $8.7\pm 1.6$ & $3.0\pm 0.3$ & $1.5\pm 0.4$ & $0.2\pm 0.1$ & $0.8\pm 0.1$ & $0.56\pm 0.08$ & $1.27\pm 0.15$ & 33650.7/32951 \\
$5$--$18$ & $9.1\pm 0.5$ & $3.05\pm 0.07$ & $1.71\pm 0.12$ & $2.7\pm 0.2$ & $7.4\pm 0.3$ & $2.46\pm 0.08$ & $1.70\pm 0.05$ & 39746.3/38544 \\
$18$--$60$ & $10.3\pm 0.2$ & $3.19\pm 0.04$ & $2.14\pm 0.08$ & $25.0\pm 0.8$ & $44.7\pm 0.9$ & $1.72\pm 0.06$ & $1.83\pm 0.03$ & 48983.6/45484 \\
$60$--$120$ & $10.0\pm 0.2$ & $4.10\pm 0.05$ & $2.75\pm 0.12$ & $39.8\pm 1.6$ & $75.3\pm 1.7$ & $1.0\pm 0.1$ & $1.73\pm 0.02$ & 52423.1/49521 \\
$120$--$180$ & $9.5\pm 0.2$ & $3.11\pm 0.03$ & $3.10\pm 0.06$ & $99.1\pm 0.9$ & $0.7\pm 0.4$ & $0.14\pm 0.04$ & $1.57\pm 0.02$ & 53470.1/51021 \\
$180$--$240$ & $9.7\pm 0.2$ & $3.35\pm 0.04$ & $3.34\pm 0.17$ & $93.1\pm 0.9$ & $<0.32$ & $0.19\pm 0.05$ & $1.16\pm 0.02$ & 53906.0/51525 \\ \hline
\multicolumn{8}{c}{{\it XMM-Newton} pn}\\ \hline
$0$--$5$ & $9.2\pm 1.6$ & $3.2\pm 0.3$ & $1.5\pm 0.4$ & $0.18\pm 0.07$ & $0.87\pm 0.09$ & $0.56\pm 0.09$ & $1.19\pm 0.13$ & 17058.5/17149 \\
$5$--$18$ & $9.0\pm 0.5$ & $3.03\pm 0.08$ & $1.67\pm 0.14$ & $2.6\pm 0.2$ & $8.1\pm 0.4$ & $2.6\pm 0.1$ & $1.48\pm 0.06$ & 20386.6/20754\\
$18$--$60$ & $10.4\pm 0.2$ & $3.20\pm 0.04$ & $2.11\pm 0.08$ & $26.7\pm 1.0$ & $51.3\pm 1.0$ & $2.28\pm 0.07$ & $1.58\pm 0.03$  & 25294.5/24836 \\
$60$--$120$ & $9.8\pm 0.1$ & $4.17\pm 0.04$ & $2.68\pm 0.08$ & $35.2\pm 1.7$ & $84.9\pm 0.4$ & $1.44\pm 0.06$ & $1.60\pm 0.01$ & 26942.3/24630 \\
$120$--$180$ & $9.2\pm 0.1$ & $3.98\pm 0.02$ & $2.95\pm 0.03$ & $106.9\pm 0.1$ & $1.5\pm 0.2$ & $0.30\pm 0.03$ & $1.41\pm 0.01$ & 26862.7/26803 \\
$180$--$240$ & $9.5\pm 0.1$ & $3.18\pm 0.02$ & $3.17\pm 0.02$ & $103.4\pm 0.1$ & $<0.20$ & $0.30\pm 0.04$ & $1.03\pm 0.01$ & 27476.0/27085\\ \hline
\multicolumn{8}{c}{{\it XMM-Newton} RGS}\\ \hline
$0$--$5$ & $12.2$\tnote{e} & $3.02\pm 0.09$ & $1.6\pm 0.8$ & $<6.73$ & $22\pm 5$ & $3.4\pm 0.6$ & $0.58\pm 0.10$ & 5303.1/5662 \\
$5$--$18$ & $12.2$ & $1.70\pm 0.03$ & $1.62\pm 0.12$ & $51\pm 2$ & $4.0\pm 0.2$ & $1.0\pm 0.1$ & $1.09\pm 0.10$ & 6799.2/6824 \\
$18$--$60$ & $12.2$ & $3.55\pm 0.08$ & $2.3\pm 0.2$ & $54\pm 4$ & $114\pm 6$ & $3.4\pm 0.4$ & $1.39\pm 0.11$ & 7685.8/7271 \\
$60$--$120$ & $12.2$ & $2.52\pm 0.11$ & $2.46\pm 0.17$ & $114\pm 4$ & $4.8\pm 0.8$ & $9.4\pm 0.1$ & $1.46\pm 0.10$ & 7585.3/7266 \\ \hline
\end{tabular}

\begin{tablenotes}
\item[a] The absorption cross sections are taken from
\citet{Verner96}.
\item[b] The VEM-weighted average temperature.
\item[c] The volume emission measure (VEM) is
given as $\int n_e n_\textrm{H} dV/(4\pi D^2)$,
where $V$ and $D$ are the volume of the emission region (cm$^3$)
and the distance to the emitting source (cm), respectively.
\item[d] We fix this parameter to the best-fitting value.
\item[e] Hydrogen column density towards the Centaurus cluster
taken from \url{http://www.swift.ac.uk/analysis/nhtot/}.
\end{tablenotes}

\end{threeparttable}
\end{table*}

\begin{table*}
\centering
\caption{Abundances of N, O, Ne, Mg, Si, S, Ar, Ca, and Fe 
derived from the local fits (CCDs) and the global fits (RGS) with respect to the proto-solar values in \citet{Lodders09}.\label{tab:parameter_2}}

\begin{tabular}{@{}c@{}c@{\:\:\:}c@{\:\:\:}c@{\:\:\:}c@{\:\:\:}c@{\:\:\:}c@{\:\:\:}c@{\:\:\:}c@{\:\:\:}c} % manual @ spacing to prevent this being too wide for a page
\hline\hline
Radial range & N & O & Ne & Mg & Si & S & Ar & Ca & Fe \\
(arcsec) & (solar) & (solar) & (solar) & (solar) & (solar) & (solar) & (solar) & (solar) & (solar) \\
\hline
\multicolumn{10}{c}{{\it Chandra} ACIS-S (local)}\\ \hline
$0$--$5$ & -- & $0.34\pm 0.08$ & $1.34\pm 0.10$ & $1.12\pm 0.05$ & $1.09\pm 0.03$ & $1.22\pm 0.07$ & $1.13\pm 0.23$ & $1.28\pm 0.35$ & $0.71\pm 0.16$ \\
$5$--$18$ & -- & $1.02\pm 0.11$ & $2.73\pm 0.17$ & $1.51\pm 0.11$ & $1.70\pm 0.06$ & $2.04\pm 0.07$ & $1.81\pm 0.15$ & $2.4\pm 0.2$ & $0.8\pm 0.2$ \\
$18$--$60$ & -- & $1.32\pm 0.05$ & $1.31\pm 0.08$ & $1.22\pm 0.04$ & $1.65\pm 0.02$ & $1.93\pm 0.03$ & $1.86\pm 0.06$ & $2.14\pm 0.08$ & $1.15\pm 0.12$ \\
$60$--$120$ & -- & $1.43\pm 0.09$ & $1.55\pm 0.08$ & $1.15\pm 0.18$ & $1.68\pm 0.03$ & $1.70\pm 0.03$ & $1.51\pm 0.06$ & $1.82\pm 0.07$ & $1.85\pm 0.19$ \\
$120$--$180$ & -- & $1.09\pm 0.10$ & $1.30\pm 0.08$ & $1.23\pm 0.17$ & $1.42\pm 0.03$ & $1.30\pm 0.04$ & $1.00\pm 0.07$ & $1.38\pm 0.08$ & $1.35\pm 0.13$ \\
$180$--$240$ & -- & $1.28\pm 0.17$ & $1.11\pm 0.10$ & $1.1\pm 0.3$ & $1.08\pm 0.04$ & $0.99\pm 0.05$ & $0.89\pm 0.10$ & $0.93\pm 0.10$ & $1.07\pm 0.13$ \\ \hline
\multicolumn{10}{c}{{\it XMM-Newton} MOS (local)}\\ \hline
$0$--$5$ & -- & $0.80\pm 0.17$ & $2.5\pm 0.3$ & $1.3\pm 0.2$ & $1.54\pm 0.14$ & $1.8\pm 0.2$ & $0.9\pm 0.5$ & $<0.60$ & $1.0\pm 0.7$ \\
$5$--$18$ & -- & $1.11\pm 0.09$ & $2.87\pm 0.12$ & $1.40\pm 0.09$ & $2.08\pm 0.06$ & $2.00\pm 0.09$ & $1.81\pm 0.19$ & $2.5\pm 0.3$ & $1.5\pm 0.3$ \\
$18$--$60$ & -- & $1.55\pm 0.06$ & $1.64\pm 0.07$ & $1.53\pm 0.04$ & $2.03\pm 0.03$ & $2.14\pm 0.04$ & $2.17\pm 0.08$ & $2.88\pm 0.10$ & $1.83\pm 0.12$ \\
$60$--$120$ & -- & $1.26\pm 0.06$ & $1.01\pm 0.05$ & $1.15\pm 0.04$ & $1.81\pm 0.02$ & $1.65\pm 0.03$ & $1.72\pm 0.07$ & $2.42\pm 0.08$ & $1.82\pm 0.08$ \\
$120$--$180$ & -- & $1.30\pm 0.07$ & $1.46\pm 0.06$ & $1.11\pm 0.05$ & $1.59\pm 0.03$ & $1.38\pm 0.04$ & $1.34\pm 0.09$ & $1.68\pm 0.08$ & $1.69\pm 0.08$ \\
$180$--$240$ & -- & $1.19\pm 0.09$ & $1.19\pm 0.07$ & $0.98\pm 0.06$ & $1.14\pm 0.03$ & $0.94\pm 0.04$ & $0.97\pm 0.11$ & $1.48\pm 0.10$ & $1.21\pm 0.06$ \\ \hline
\multicolumn{10}{c}{{\it XMM-Newton} pn (local)}\\ \hline
$0$--$5$ & -- & $0.76\pm 0.13$ & $2.2\pm 0.4$ & $1.26\pm 0.17$ & $1.7\pm 0.2$ & $1.7\pm 0.2$ & $0.8\pm 0.6$ & $1.2\pm 0.7$ & $0.6\pm 0.4$ \\
$5$--$18$ & -- & $0.77\pm 0.07$ & $2.60\pm 0.12$ & $1.41\pm 0.10$ & $1.65\pm 0.06$ & $1.85\pm 0.09$ & $1.40\pm 0.19$ & $1.6\pm 0.3$ & $1.8\pm 0.3$ \\
$18$--$60$ & -- & $0.85\pm 0.04$ & $1.22\pm 0.07$ & $1.30\pm 0.05$ & $1.65\pm 0.03$ & $1.75\pm 0.04$ & $1.60\pm 0.08$ & $2.09\pm 0.10$ & $1.87\pm 0.10$ \\
$60$--$120$ & -- & $0.69\pm 0.05$ & $1.22\pm 0.07$ & $1.15\pm 0.05$ & $1.55\pm 0.03$ & $1.46\pm 0.03$ & $1.43\pm 0.07$ & $1.61\pm 0.08$ & $1.77\pm 0.06$ \\
$120$--$180$ & -- & $0.72\pm 0.08$ & $1.71\pm 0.07$ & $1.07\pm 0.06$ & $1.31\pm 0.04$ & $1.16\pm 0.05$ & $1.04\pm 0.10$ & $1.25\pm 0.09$ & $1.45\pm 0.06$ \\
$180$--$240$ & -- & $0.51\pm 0.07$ & $1.54\pm 0.11$ & $0.84\pm 0.10$ & $0.95\pm 0.03$ & $0.88\pm 0.05$ & $0.81\pm 0.12$ & $0.85\pm 0.10$ & $1.00\pm 0.04$ \\ \hline
\multicolumn{10}{c}{{\it XMM-Newton} RGS (global)}\\ \hline
$0$--$5$ & $0.8\pm 0.4$ & $0.54\pm 0.09$ & $0.42\pm 0.10$ & $0.42\pm 0.09$ & -- & -- & -- & -- & $0.58\pm 0.10$ \\
$5$--$18$ & $2.0\pm 0.5$ & $0.93\pm 0.09$ & $0.74\pm 0.10$ & $0.82\pm 0.12$ & -- & -- & -- & -- & $1.09\pm 0.10$ \\
$18$--$60$ & $2.3\pm 0.5$ & $1.31\pm 0.11$ & $0.99\pm 0.14$ & $1.03\pm 0.12$ & -- & -- & -- & -- & $1.39\pm 0.11$ \\
$60$--$120$ & $<0.47$ & $1.40\pm 0.15$ & $0.76\pm 0.15$ & $0.62\pm 0.18$ & -- & -- & -- & -- & $1.46\pm 0.10$ \\ \hline
\end{tabular}
\end{table*}

\begin{table}
\centering
\caption{Abundances of Fe-peak elements derived from the local fits (CCDs, except for Ni of ACIS)
and global fits (Ni of ACIS and RGS) with respect to the proto-solar values in \citet{Lodders09}.\label{tab:parameter_3}}

\begin{tabular}{@{}c@{}c@{\:\:\:}c@{\:\:\:}c} % manual @ spacing to prevent this being too wide for a page
\hline\hline
Radial range & Cr & Mn & Ni \\
(arcsec) & (solar) & (solar) & (solar) \\
\hline
\multicolumn{4}{c}{{\it Chandra} ACIS-S}\\ \hline
$0$--$5$ & $<2.69$ & $<6.43$ & $3.5\pm 0.4$ \\
$5$--$18$ & $<1.52$ & $<3.31$ & $4\pm 3$ \\
$18$--$60$ & $2.0\pm 0.6$ & $<1.30$ & $3.13\pm 0.10$ \\
$60$--$120$ & $1.6\pm 0.5$ & $2.2\pm 1.1$ & $2.40\pm 0.07$ \\
$120$--$180$ & $<0.30$ & $<0.75$ & $2.02\pm 0.08$ \\
$180$--$240$ & $0.8\pm 0.6$ & $<0.48$ & $1.61\pm 0.11$ \\ \hline
\multicolumn{4}{c}{{\it XMM-Newton} MOS}\\ \hline
$0$--$5$ & $<4.67$ & $<3.44$ & $<15$ \\
$5$--$18$ & $<1.70$ & $<2.55$ & $<2.11$ \\
$18$--$60$ & $2.6\pm 0.7$ & $2.0\pm 1.0$ & $3.4\pm 0.8$ \\
$60$--$120$ & $2.0\pm 0.5$ & $2.3\pm 0.8$ & $3.2\pm 0.5$ \\
$120$--$180$ & $1.5\pm 0.5$ & $1.3\pm 0.9$ & $1.9\pm 0.6$ \\
$180$--$240$ & $0.7\pm 0.5$ & $1.0\pm 0.9$ & $1.0\pm 0.5$ \\ \hline
\multicolumn{4}{c}{{\it XMM-Newton} pn}\\ \hline
$0$--$5$ & $<2.38$ & $<3.04$ & $<4.16$ \\
 $5$--$18$ & $3.7\pm 1.7$ & $5\pm 3$ & $<1.99$ \\
 $18$--$60$ & $2.7\pm 0.6$ & $3.6\pm 1.0$ & $3.5\pm 0.7$ \\
 $60$--$120$ & $1.6\pm 0.4$ & $1.2\pm 0.7$ & $3.7\pm 0.5$ \\
 $120$--$180$ & $1.5\pm 0.5$ & $2.1\pm 0.7$ & $2.3\pm 0.5$ \\
 $180$--$240$ & $1.3\pm 0.5$ & $1.4\pm 0.7$ & $1.8\pm 0.4$ \\ \hline
\multicolumn{4}{c}{{\it XMM-Newton} RGS}\\ \hline
$0$--$5$ & -- & -- & $2.2\pm 0.4$ \\
$5$--$18$ & -- & -- & $2.5\pm 0.3$ \\
$18$--$60$ & -- & -- & $2.1\pm 0.2$ \\
$60$--$120$ & -- & -- & $1.2\pm 0.3$ \\ \hline
\end{tabular}
\end{table}

\subsubsection{RGS spectra}
\label{subsubsec:rgs_fit}

\begin{figure*}
\includegraphics[width=\textwidth]{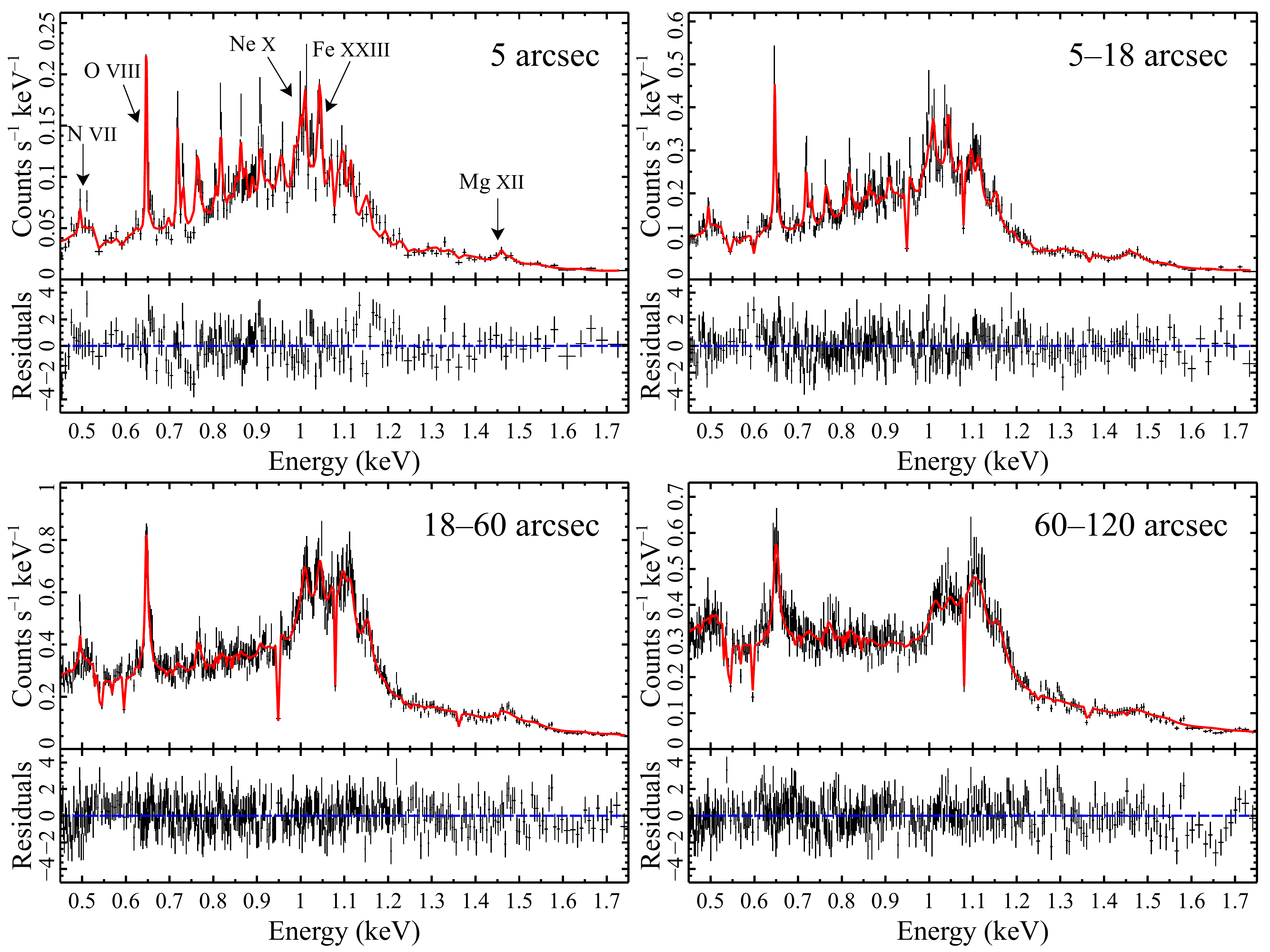}
\caption{The RGS spectra extracted from each cross-dispersion slice.
The black points indicate the data from the observation in 2007
and the red lines are the best-fitting model summarised in
Table\,\ref{tab:parameter_1} and \ref{tab:parameter_2},
where only the first-order spectra are shown for clarity.
Several emission line structures are marked on top left panel.
\label{fig:rgs_spec}}
\end{figure*}

We fit the high-resolution X-ray spectra of RGS
with the almost same spectral model described in \S\ref{subsubsec:ccd_fit},
considering the spectral broadening effect due to
the spatial extent of the source using {\tt rgsxsrc} model with MOS1 images.
The first- and second-order spectra
for each observation are simultaneously fitted.
Since contributions of the astrophysical background emissions are relatively minor
to the spectra of the innermost core regions,
we only take into account the instrumental component
with a steep power-law model.
We assume the line and continuum temperatures are the same because the 
continuum temperatures are poorly constrained with the RGS spectra.
We also fix the Galactic absorption to the values estimated
through the tool of \citet{Willingale13}.
As shown in Fig.\,\ref{fig:rgs_spec},  
RGS enables us to rather resolve the Ne lines from the Fe-L lines.
In addition, N-Ly~$\alpha$ emission line ($\sim$\,0.5\,keV) is clearly seen in RGS spectra.
Then, we allow N, O, Ne, Mg, Fe and Ni abundances to vary, and C abundance is fixed to the solar value.
The other element abundances are tied to those of the nearest lower atomic number element,
which is allowed to vary.
As shown in Fig.\,\ref{fig:rgs_spec}, 
this model yields good fits with C-stat/dof $\sim$\,1.0.
The best-fitting parameters are summarised in Table\,\ref{tab:parameter_1}, \ref{tab:parameter_2},
and Fig.\,\ref{fig:rgs_spec}, respectively.
Although the individual spectral extraction regions for RGS are not identical to those
for the CCDs as indicated in \S\,\ref{subsec:xmm_reduction},
the VEM-weighted average temperatures from the two kinds of detectors are consistent
when the inner radius of the CCD region and the inner cross-dispersion 
distance of the RGS are the same (Fig.\,\ref{fig:kT_prof}).

\subsection{Abundance Profiles}
\label{subsec:profile}

\begin{figure*}
\includegraphics[width=\textwidth]{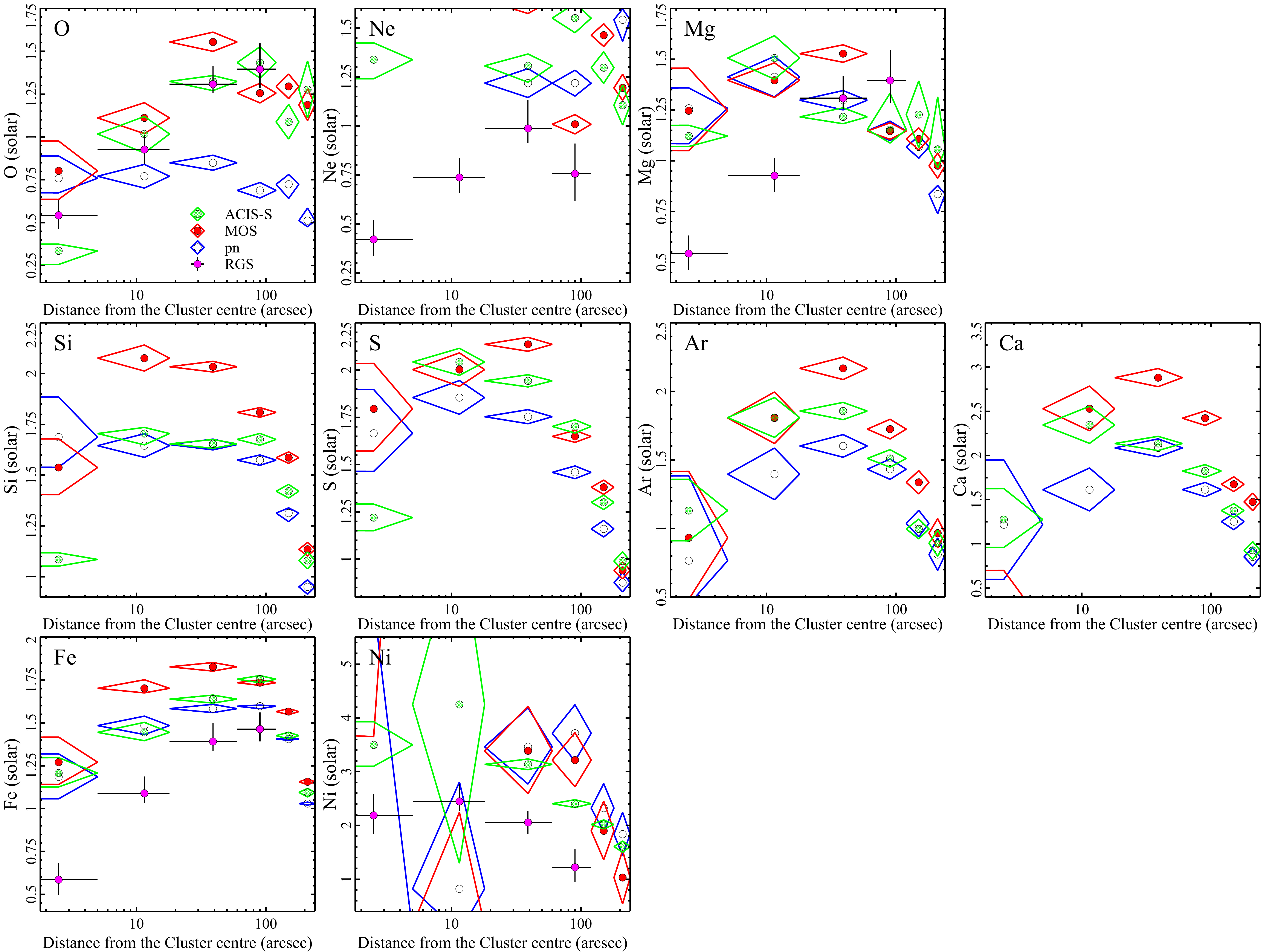}
\caption{Radial profiles of the metal abundances
(O, Ne, Mg, Si, S, Ar, Ca, and Fe) obtained using a triple CIE model.
The green-, red-, and blue-diamond plots are for results with {\it Chandra} ACIS-S, {\it XMM-Newton} MOS,
pn, respectively. The magenta plots indicate the RGS results of O, Ne, Mg, Fe, and Ni.
We plot the Ne abundance setting the y-axis range according to values with RGS for clarity.
\label{fig:abund_profile}}
\end{figure*}

As described in Section\,\ref{sec:intro}, our purpose is to study
the central metal distribution in the Centaurus cluster.
In Fig.\,\ref{fig:abund_profile}, we show radial profiles of elemental abundances
except for those of Cr and Mn, which have large statistical uncertainties.
The derived abundances of Fe and IMEs show steep negative gradients
beyond 120\,arcsec from the cluster centre, 
are peaked at 60--120\,arcsec and abrupt drops within the central 18\,arcsec.
In contrast, O, Ne and Mg have somewhat flatter profiles beyond 120\,arcsec.
The discrepancies among the abundances from different detectors are significant,
about a factor of two at most, especially within the Fe abundance peak.
The RGS tends to give lower values of metal abundances than those from the CCDs,
with a more significant abundance drop than that with CCDs.
These profiles obtained from our analysis are roughly consistent with
those from previous works \citep[][]{Panagoulia13,Sanders16,Liu19b,Lakhchaura19}.
Moreover, our RGS analysis provides first the Ne abundance profile
showing a similar drop with Fe, although there had been no hint of declination
of the Ne abundance from works with CCD data \citep[][]{Liu19b}.

\begin{figure*}
\includegraphics[width=\textwidth]{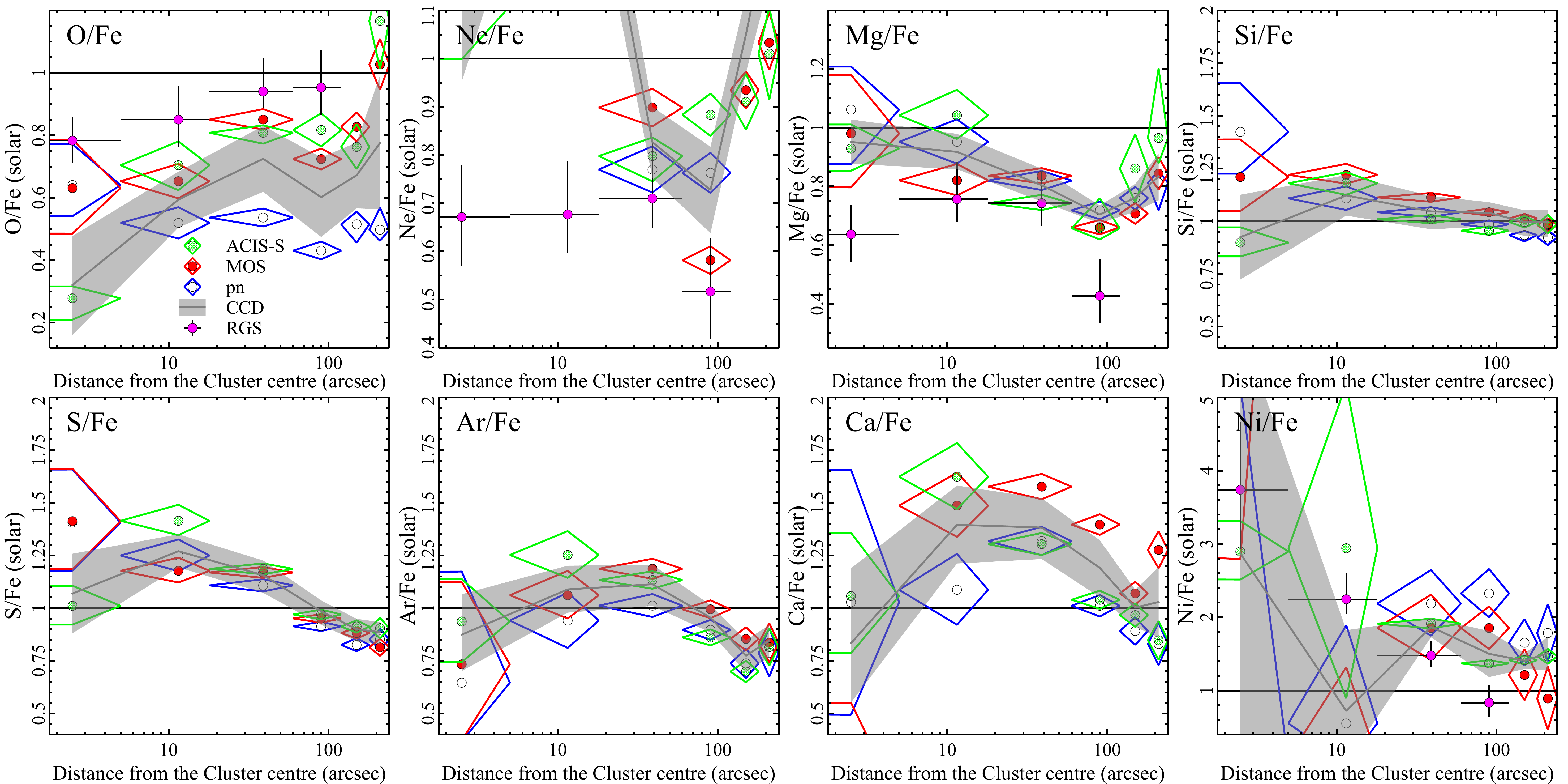}
\caption{Radial profiles of the relative abundance ratios to Fe
obtained using a triple CIE model. The diamonds follow the same manner
as Fig\,\ref{fig:abund_profile}.
The shaded areas indicate the mean values among three CCD detectors.
We also show the solar ratio with solid lines on each panel.
\label{fig:relabund_profile}}
\end{figure*}

As shown in Fig.\,\ref{fig:relabund_profile}, 
the abundance ratios to Fe, X/Fe,  show  flat distributions, although 
the S/Fe,  Ar/Fe, and Ca/Fe ratios within the central 60\,arcsec are higher
by a few tens of per cent than those beyond 120\,arcsec.
The scatters in the X/Fe ratios among
the different detectors are much smaller than those in the absolute abundances.
Exceptionally,  RGS and CCDs yield significantly different Ne/Fe ratios.
Considering that the Ne-K lines are blended in the Fe-L lines with the CCD
as mentioned in \S\,\ref{subsubsec:ccd_fit},
we expect RGS gives more reliable Ne/Fe ratios.
Although the O/Fe ratios with pn are about a factor of two lower than those with MOS,
which has a slightly better energy resolution than pn, are consistent with those derived from the RGS spectra.
Therefore, for CCSN products of O, Ne, and additionally Mg,
we will discuss only the results with RGS data,
which would provide a better estimate of these abundances with its high spectral resolution.
The Ca/Fe ratios among the CCD detectors show discrepancies of about 50 per cent at most, while
the other IMEs/Fe ratios are consistent with each other.

\subsection{Systematic Uncertainties}
\label{subsec:sys_err}

Various systematic uncertainties sometimes give biases in abundance measurements
in cool cores \citep[e.g.,][]{Werner08}.
For example,  simple temperature modelling to a complex structure in cores
would lead to underestimation of abundances, especially when fitting spectra with
strong Fe-L lines \citep[e.g.,][]{Buote97,Matsushita03}. 
Therefore, we also estimate the abundances using other
temperature structure models; two and four CIE components.
For the four-CIE model, the temperature of the coolest component
is half of that of the second coolest one as the temperatures of this model making
a geometric sequence with a common ratio of 0.5.
Even for the innermost region, these models give almost the same Fe abundances within several per cent
with those derived in the previous subsection.
Since the quadruple model yields  unnaturally
low VEM for one component ($<$\,1 per cent of the other three) and almost the same C-statistic values,
we conclude that the three temperature-components model is sufficient 
to model the emission from the Centaurus core.

Projection of X-ray emission along the line of sight
could also obscure the actual distribution of temperature
or abundances.  Since the Centaurus cluster core deviates from spherical
symmetry due to the plume-like structure,  we do not apply the deprojection method
often used to analyse the X-ray emission from cluster cores
\citep[e.g.,][]{Ettori02,Ikebe04,Russel08}.
Instead, we fit the innermost spectra using
a sextuple-CIE model. Here, we first fit the spectra of the outermost bin,
or 180--240\,arcsec region,
with a single-CIE model. Then, we fit the spectra of 120--180\,arcsec region
with a two-CIE model, where one CIE component has the same temperature
and metal abundances as those at the outermost region.
Finally, we fit the innermost spectra using a  sextuple-CIE model.
Then, we get almost the same metal abundances within 10 per cent as those
with the three-component CIE model.

\begin{figure*}
\includegraphics[width=\textwidth]{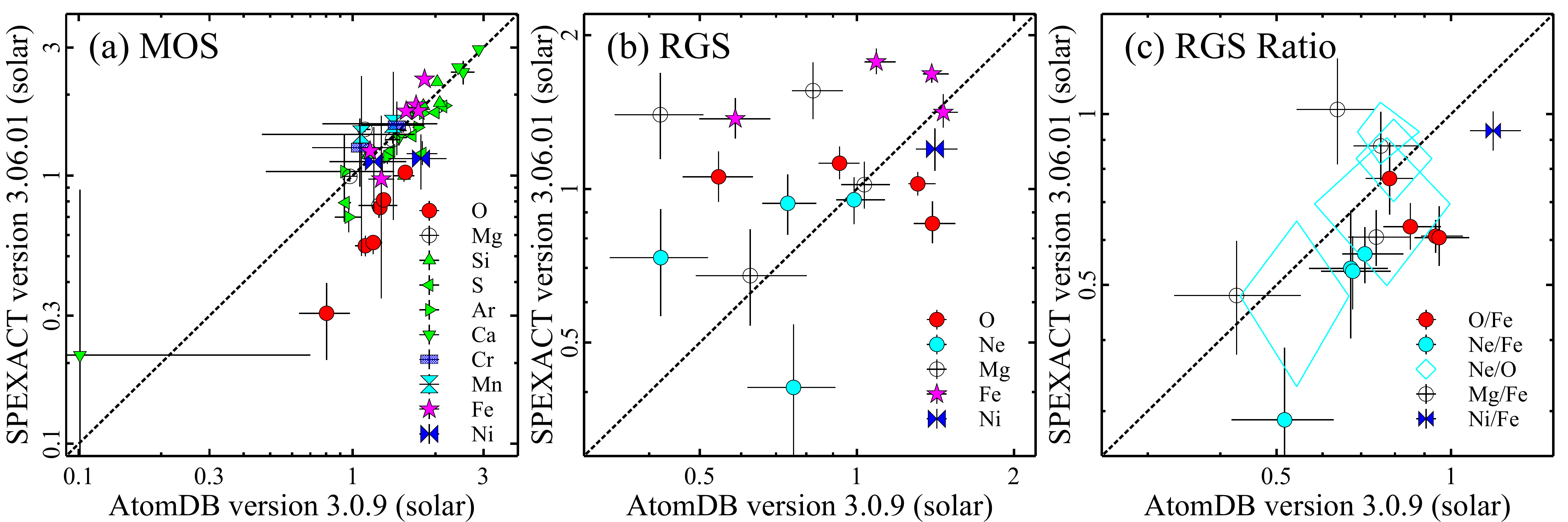}
\caption{(a) A comparison of the metal abundances
(O, Mg, Si, S, Ar, Ca, Fe, Cr, Mn, and Ni) in each annular region of the MOS data
with AtomDB version 3.0.9 and SPEXACT version 3.06.01.
The Ne abundance is absent in the plots
for its unreliability (see \S\,\ref{subsec:profile}).
For the Cr, Mn, and Ni abundance, we show the averaged values on
the central $<$\,120\,arcsec core and the outer 120--240\,arcsec region.
(b) Same as (a) for O, Ne, Mg, Fe, and Ni taken from the RGS data.
(c) A comparison of the relative abundance ratios
for O/Fe, Ne/Fe, Ne/O, Mg/Fe, and Ni/Fe obtained from the RGS data.
The dotted lines shown in each panel indicate the equal value between two codes.
\label{fig:atomcode}}
\end{figure*}

Next, in order to study the uncertainties caused by systematics in 
the atomic data, we fit the spectra of the MOS and RGS 
with the SPEXACT version 3.06.01 \citep{Kaastra96}.
We plot the derived abundances and relative abundance ratios to Fe
with AtomDB and SPEXACT in Fig.\,\ref{fig:atomcode}.
We do not report the results for Ne in Fig.\,\ref{fig:atomcode}(a)
because the Ne-K lines are blended into the Fe-L complex in CCD spectra.
With MOS data, the latest version of SPEXACT gives almost the same abundances
except for the O abundances.
With RGS, SPEXACT yields nearly constant metal abundances in the core,
which are significantly different from those with AtomDB (Fig.\,\ref{fig:atomcode}(b)).
On the other hand, the abundance ratios with RGS are mostly consistent
within a few tens of per cent
for all elements between both atomic codes (Fig.\,\ref{fig:atomcode}(c)).
The differences of $\sim$\,20 per cent in the O/Fe ratios and Ne/Fe ratios
between AtomDB and SPEXACT
may be caused by systematic uncertainties in the Fe-L atomic code.
We note that the Ne/O ratios with RGS from SPEXACT and AtomDB agree with each other.

\section{DISCUSSION}
\label{sec:discussion}

\subsection{Abundance Measurements in the Innermost Regions}
\label{subsec:drop_origin}

As described in Section\,\ref{sec:intro}, 
the mechanisms of the abundance depletions sometimes reported in cluster cores are still under discussion.
We also detect abrupt abundance drops within central 18\,arcsec for most elements.
\citet{Panagoulia13} and \citet{Lakhchaura19} proposed that these drops arise from cool dust grains
in the cluster centre to which a significant fraction of metals is locked up.
If this is the case, we expect that the abundance of 
non-reactive elements, i.e., the noble gas, show no central drop.
\citet{Lakhchaura19} reported that only the Ar/Fe ratio
in the Centaurus cluster core slightly increases towards the centre
while the Ar abundance still shows a central drop as other metals.
They concluded that the drops are due to the incorporation of metals into dust grains.
Our Ar/Fe profile from the longer exposure data also shows a slight increase
towards the cluster centre from 120\,arcsec and is consistent within error bars
with that derived by \citet{Lakhchaura19}.
However, we get a flat profile within $\lesssim$\,60\,arcsec
where the abundance drop occurs (Fig.\,\ref{fig:relabund_profile}).
We note that the increasing profile of the Ar/Fe ratio by \citet{Lakhchaura19}
is partially attributed to their region selection with the $\lesssim$\,30\,arcsec innermost bin.
This larger radius than ours is suitable for {\it XMM-Newton} but probably makes a too much concession
to the extremely high angular resolution of the {\it Chandra} ACIS ($\sim$\,0.5\,arcsec),
which could blur the innermost metal distribution in exchange for photon statistics.
Our Ar/Fe profile is quite similar to that of the Ca/Fe ratios
even though Ca can be easily trapped into dust grains.
In addition, we find that the Ne/Fe profile
using RGS is also flat towards the centre.
These results do not support the grain origin for the abundance drops.

\begin{figure*}
\includegraphics[width=\textwidth]{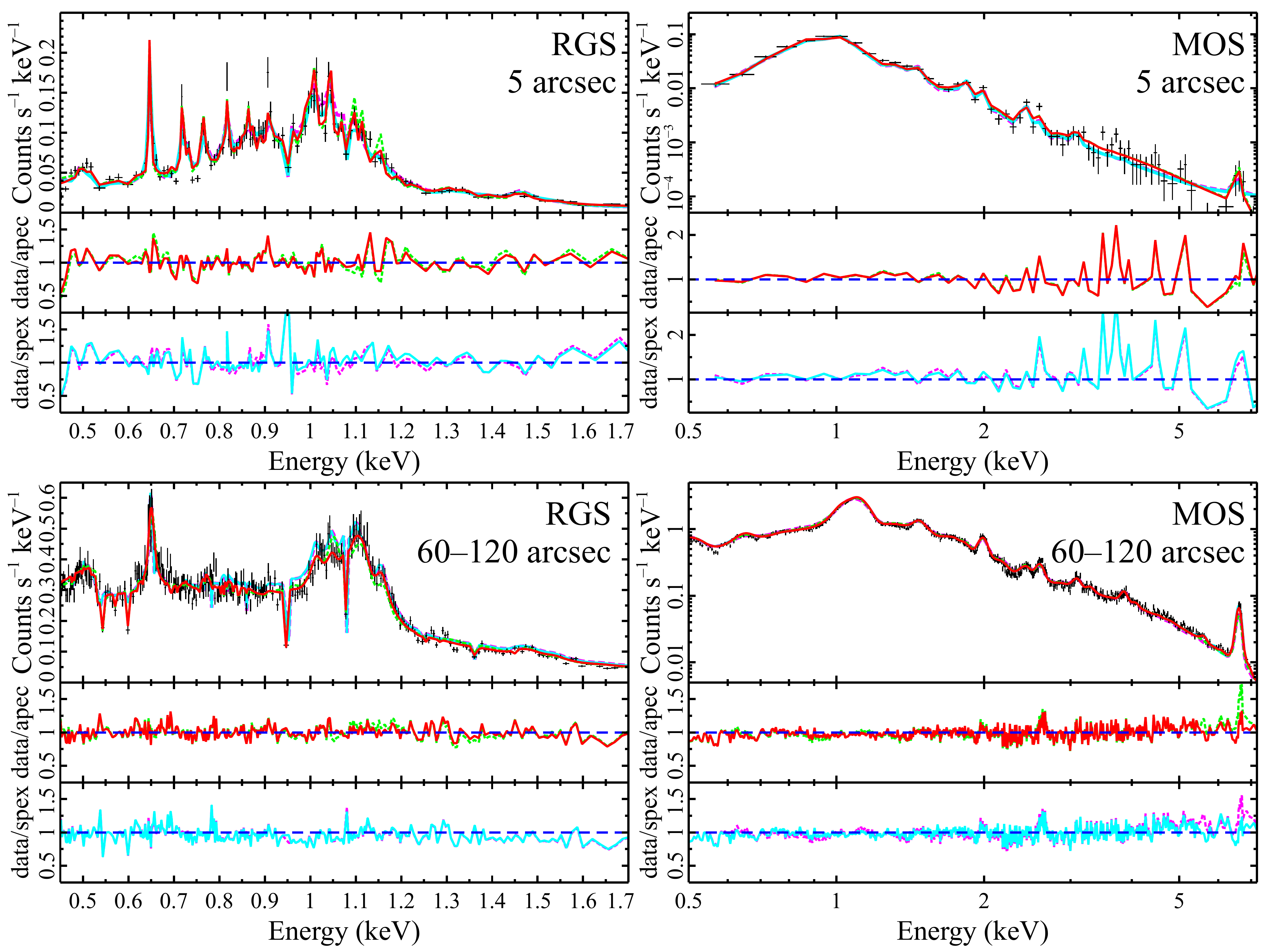}
\caption{A comparison of two models and atomic codes.
The spectra at the innermost region and  at the peak
of the Fe abundance are plotted for RGS and MOS.
The red solid line represents the best-fitting model
(Fe $\sim$\,0.5\,solar at $<$\,5\,arcsec,
Fe $\sim$\,1.7\,solar at 60--120\,arcsec)
with AtomDB and the green dotted one is for the fixed Fe abundance model
(Fe $= 1.7$\,solar at $<$ 5\,arcsec, Fe $= 0.5$\,solar
at 60--120\,arcsec), respectively.
The cyan solid and magenta dotted lines follow
the same plotting manner but using SPEXACT, where the best-fitting and
peak Fe abundances are 1.4 and 1.7\,solar, respectively.
We also plot the divided data by each model, where data/apec is
for the results with AtomDB and data/spex is for SPEXACT.
\label{fig:rgs_nodrop}}
\end{figure*}

The different detectors and different plasma codes give
significantly inconsistent elemental abundances, while
the discrepancies in the abundance ratios are relatively small.
To investigate the origin of these discrepancies, we fix the Fe abundance
at the peak value ($\sim$\,1.7\,solar at 60--120\,arcsec) and
fit the spectra of the innermost region ($<$\,5\,arcsec)
of RGS and MOS.  As shown In Fig.\,\ref{fig:rgs_nodrop},
this high Fe abundance model reproduces the continuum and most of the line emissions
for both AtomDB and SPEXACT, yielding similar abundance ratios.
With AtomDB, the derived  C-stat/dof for the RGS spectrum, 5363.8/5663, is
significantly larger than 5303.1/5662 with the best-fitting Fe abundance at 0.58\,solar.
This large C-statistic is mainly caused by residual structures at 1.1--1.2\,keV energy band,
rather than the difference in the equivalent width of lines.
In contrast,  as shown In Fig.\,\ref{fig:rgs_nodrop}, SPEXACT,
which yields the best-fitting Fe abundance of 1.4\,solar at the innermost region,
does not show such residual structures.
When we exclude the 0.9--1.2\,keV band where the abovementioned discrepancy
between two codes is exhibiting, the derived Fe abundance with AtomDB becomes 1.2$\pm$0.3\,solar.
This value mostly converges to that with SPEXACT and also that with MOS data,
although the Fe abundance with MOS data under the same excision is not significantly changed.
Thus, while the Fe abundance drop is not dissolved completely,
quite a sharp drop with RGS data could be attenuated.
The disagreement between the latest version of AtomDB and SPEXACT 
occurs for \ion{Fe}{XXIII} and \ion{Ne}{X} emission lines around 1.2\,keV.
These deficits around this energy band 
have been reported since {\it ASCA}, especially in luminous X-ray sources
like binaries or SN remnants
\citep[e.g.,][]{Brickhouse00,Katsuda15,Nakano17}.
We also fit the RGS and MOS spectra at the 60--120\,arcsec region in the same way,
fixing the Fe abundance at 0.5 and 1.7\,solar.
Then, as seen in the bottom-right panel of Fig.\,\ref{fig:rgs_nodrop},
the low Fe abundance models cannot reproduce the line strength of the Fe line at 6.7\,keV.

The systematic uncertainties in the derived abundances from $\sim$\,1\,keV plasma
have been discussed \citep[e.g.,][]{Arimoto97,Matsushita97,Matsushita00,Mernier20a,Gastaldello21}.
The two-photon and free-bound emissions by heavy elements contribute to the continuum, 
especially for the lower temperature plasma in the innermost regions. 
Spectral fits try to reduce C-statistics for the lower energy band with high statistics,
where the line emissions dominate the continuum.
Then, minor systematic uncertainties in the atomic data and/or in the response matrices
can bias the derived abundances.
The local fits around the Fe-He~$\alpha$ line also give abundance drops,
where the atomic data is expected to be much more reliable than those for Fe-L lines.
This is because the abundances of the other elements are fixed to the best-fitting values from the global fits.
For example, when we fix the $\alpha$-elements abundances at 1.3\,solar,
the local fits for the {\it Chandra} $<5$\,arcsec region gives the Fe abundance of 1.1\,solar.
Thus,  measurements of absolute abundances are challenging 
especially in the very core gas, where the abundance drops are reported.
In the hot outer regions in which the prominent Fe-K line is detected,
the absolute abundances thereof are more reliable.
In contrast, the abundance ratios are determined from the ratios of line strengths.
As a result, the two atomic codes and different CCDs yield more consistent abundance ratios
than the absolute abundances.

\subsection{Abundance pattern of the Centaurus cluster core}

\begin{figure*}
\includegraphics[width=\textwidth]{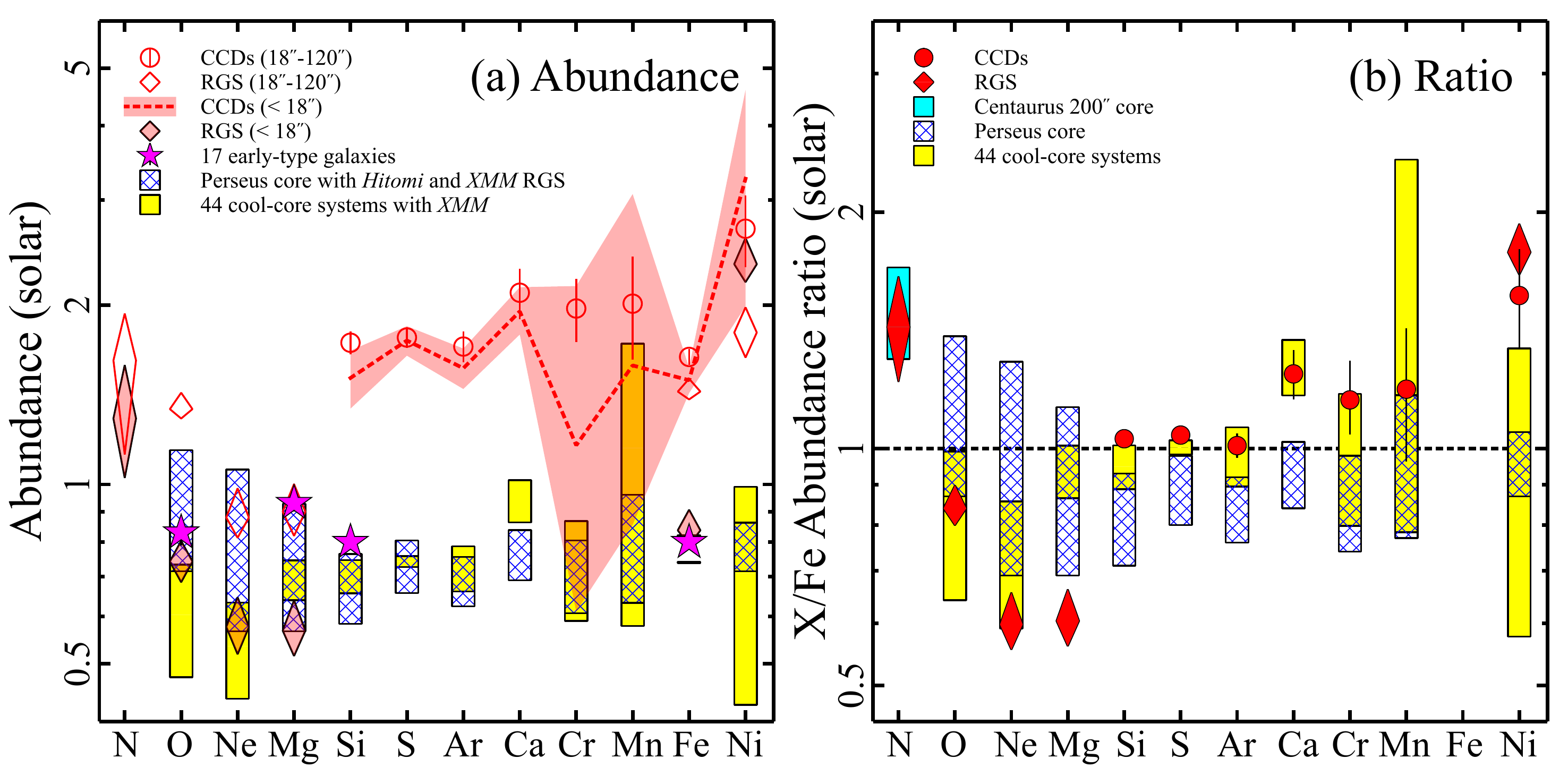}
\caption{(a) Abundance pattern measured in $<$\,18\,arcsec
and 18--120\,arcsec region of the Centaurus cluster.
The open circle and diamond plots show the results
of 18--120\,arcsec region derived from CCDs and RGS, respectively.
The shaded area and filled diamonds are from the central 18\,arcsec core
with CCDs and RGS, respectively.
For comparison, we plot the patterns in the ICM from other works,
where the star plots are from 17 early-type galaxies \citep{Konami14},
the blue boxes are from the Perseus cluster \citep{Simionescu19},
and the yellow boxes are from 44 cool-core clusters, groups,
and ellipticals \citep[CHEERS sample;][]{Mernier18b}.
(b) Abundance ratio pattern obtained from the central
120\,arcsec core in the same manner of the left.
The cyan box indicates the N/Fe ratio in the Centaurus cluster from \citet{Mao19},
while the blue and yellow boxes are from \citet{Simionescu19} and \citet{Mernier18b}.
\label{fig:abund_pattern}}
\end{figure*}

We plot the absolute and relative abundance pattern in the ICM of the Centaurus core
(Fig.\,\ref{fig:abund_pattern}(a) and (b)).
Here we use the weighted averages of the derived abundances within ($<$\,18\,arcsec)
and outside (18--120\,arcsec) the abundance drop.
Outside the abundance drop, the metal abundances from N to Ni  are
over-solar values, consistent with previous X-ray studies
\citep[e.g.,][]{Matsushita07,Takahashi09,Sakuma11,Sanders16}.
These abundances are significantly larger than those
observed in the Perseus cluster core with {\it Hitomi} \citep[][]{Simionescu19},
other cool-core systems including groups \citep[][]{Mernier18b},
and early-type galaxies \citep[][]{Konami14}.
The statistical uncertainties in Cr, Mn, and Ni abundances are comparable to
those of the Perseus cluster core \citep[][]{Mernier18b,Simionescu19}.
As shown in Fig.\,\ref{fig:abund_pattern}(b), the abundance ratio
pattern in the Centaurus core is mostly consistent with  the solar composition
and those of other cool-core systems  and early-type galaxies
\citep[e.g.,][]{Mernier18b,Simionescu19, Konami14}.
The exceptions are the high Ni/Fe ratio, about 1.5--2\,solar, with RGS and CCDs,
high N/O ratio at 1.5\,solar, and low Ne/Fe and Mg/Fe ratios at 0.6\,solar with RGS.

% SN yields

\begin{figure*}
\includegraphics[width=\textwidth]{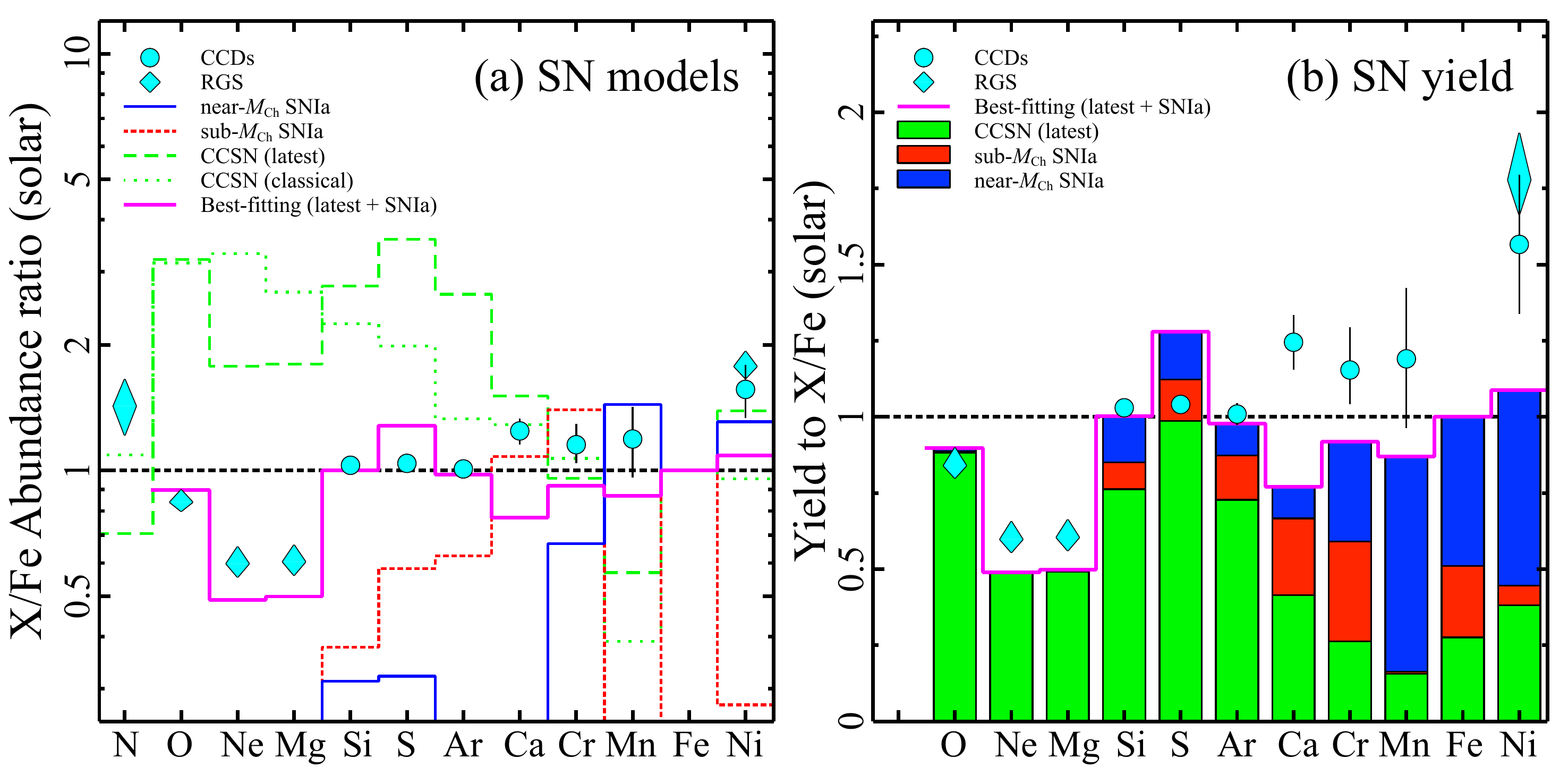}
\caption{(a) Comparison of the abundance ratio pattern same as Fig.\,\ref{fig:abund_pattern}(b)
with SN nucleosynthesis models.
The thin solid, dotted, dashed and dot-dashed lines represent the relative
abundance ratios to Fe for near-$M_\textrm{Ch}$ SNIa \citep{LN18},
for sub-$M_\textrm{Ch}$ SNIa \citep{Shen18},
for latest CCSN \citep{Sukhbold16}, and `classical' CCSN \citep{Nomoto06}, respectively.
The magenta thick line corresponds to the best-fitting model
which assumes a linear combination of SNIa and latest CCSN model.
(b) The contribution of each type of SNe to the best-fitting ratio pattern.
\label{fig:abund_fit}}
\end{figure*}

\begin{table*}
\centering
\caption{Best-fitting combination of CCSN and SNIa yield models by number to the abundance ratio pattern
observed in the core of the Centaurus cluster.\label{tab:sn_yield}}
\begin{tabular}{cccc} % manual @ spacing to prevent this being too wide for a page
\hline\hline
SN type & Model & All elements & Excluding Ne, Mg, and Ni \\
 & (Reference) & & \\
\hline
near-$M_\textrm{Ch}$ SNIa & W7 & $0.10\pm 0.03$ & $0.08\pm 0.04$ \\
 & \citep{LN18} & & \\
sub-$M_\textrm{Ch}$ SNIa & double detonation & $0.06\pm 0.04$ & $0.11\pm 0.06$ \\
 & \citep{Shen18} & & \\
CCSN & N20 & $0.84\pm 0.03$ & $0.82\pm 0.04$ \\
 & \citep{Sukhbold16} & & \\ \hline
\multicolumn{2}{c}{$\chi ^2$/dof} & 11.2/8 & 7.2/5 \\ \hline
near-$M_\textrm{Ch}$ SNIa & W7 & $0.03 \pm 0.02$ & $0.02 \pm 0.01$ \\
sub-$M_\textrm{Ch}$ SNIa & double detonation & $0.04\pm 0.02$ & $0.04\pm 0.02$ \\
CCSN & classical & $0.93\pm 0.02$ & $0.94\pm 0.02$ \\
 & \citep{Nomoto06} & & \\ \hline
\multicolumn{2}{c}{$\chi ^2$/dof} & 17.8/8 & 9.9/5 \\ \hline
\end{tabular}
\end{table*}

\subsubsection{Contributions from different SN yield models}

We consider a linear combination of recent SN nucleosynthesis calculations
to reproduce the abundance ratio pattern from the O/Fe to Ni/Fe observed in the Centaurus core.
We adopt the latest update of W7 model from 2D hydrodynamics simulations
(\citealt{LN18}, for near-$M_\textrm{Ch}$ SNIa) and
a double-degenerated double-detonation explosion of a solar mass
and solar metallicity white dwarf (\citealt{Shen18}, for sub-$M_\textrm{Ch}$ SNIa).
A recent calculation of exploding massive stars averaged over Salpeter initial mass function \citep[][]{Salpeter55}
calibrated using observed energy of SN 1987A (N20 model, \citealt{Sukhbold16})
and the classical calculations of \citet{Nomoto06} are considered for CCSN yield models.
These models are plotted in Fig.\,\ref{fig:abund_fit}(a) with the observed abundance ratio pattern.
To fit the pattern and calculate $\chi^2$, we consider not only the statistical errors of
the observed abundance ratios but also 20 per cent systematic uncertainties reflecting the discrepancies
between the derived abundance ratios from AtomDB and SPEXACT.
Table\,\ref{tab:sn_yield} summarises the fitting results with the contributions of
each SN population to the total number of SNe.
The N20 model by \citet{Sukhbold16}  gives a smaller $\chi^2$ than the `classical' model by \citet{Nomoto06}.
Since the Ne/Fe, Mg/Fe, and Ni/Fe ratios deviate from the best-fitting pattern even using the N20 model,
we excluded Ne, Mg, and Ni and fit the abundance pattern in the same way.
Then, we get a better $\chi^2$, but the contributions of each SN model are consistent 
within the error bars.
The ratio of numbers of CCSN to the total number of SNe is about 80 per cent with
the N20 model, while it is about 90 per cent with the `classical' model.
This result is consistent with those for the other clusters \citep[e.g.,][]{Simionescu15,Simionescu19}.
Fig.\,\ref{fig:abund_fit}(b) shows the contributions of CCSN,
sub-$M_\textrm{Ch}$, and near-$M_\textrm{Ch}$ SNeIa to each element
for the best-fitting combination with the N20 model.

\subsubsection{Abundance pattern of $\alpha$-elements}

Fig.\,\ref{fig:abund_fit}(a) shows that
O, Ne, and Mg are predominately synthesised by CCSNe.
The discrepancies between the observed O, Ne, and Mg abundance pattern
and those by CCSN models significantly contribute to the $\chi^2$ in Table\,\ref{tab:sn_yield}.
The observed O, Ne, and Mg abundance pattern, lower Ne and Mg abundances
compared to the O abundance,
resembles that of the latest CCSN/N20 model by \citet{Sukhbold16}
rather than the `classical' CCSN model by \citet{Nomoto06}. 
Fig.\,\ref{fig:abund_fit}(b) shows that Si, S, and Ar
are mostly synthesised by CCSNe.
The combination with the N20 model, near- and sub-$M_\textrm{Ch}$ roughly reproduces
the observed abundance pattern of these elements and Fe.
The contribution of SNeIa products is higher to Ca than those for the other IMEs,
although the observed Ca/Fe ratio is significantly higher than the best-fitting combination.

\subsubsection{Abundance pattern of Fe-peak elements}

The observed Cr/Fe and Mn/Fe pattern is explained by a combination of CCSNe, near-$M_\textrm{Ch}$ SNeIa,
and sub-$M_\textrm{Ch}$ SNeIa, as for the Perseus cluster core.
As summarised in Section\,\ref{sec:intro}, the Fe-peak elements
are mainly forged by SNeIa, especially in the hottest layers
of the exploding white dwarf \citep[e.g.,][]{Seitenzahl13}.
In particular, $^{55}$Mn is an important element since it is synthesised more by SNeIa
than CCSNe relative to Fe \citep[][]{KN09}, while Cr/Fe and Ni/Fe ratios are less independent
of the SNIa/CCSN contributions.
As shown in Fig.\,\ref{fig:abund_fit}(a),
the CCSN models predict the Mn/Fe ratios of approximately a half of the solar ratio.
The observed solar Mn/Fe ratio in the Centaurus core and other systems indicate
a significant contribution of SNeIa products to the Fe-peak elements.
The latest hydrodynamic simulations \citep[e.g.,][]{LN18}
predict that the extremely high-density core of
near-$M_\textrm{Ch}$ SNeIa synthesises a higher fraction of
neutronised isotopes like $^{55}$Mn and $^{58}$Ni
than sub-$M_\textrm{Ch}$ ones, which is also confirmed by
observational studies on SNIa remnants \citep[e.g.,][]{Yamaguchi15,Ohshiro21}.
The observed solar Mn/Fe ratio indicates a significant contribution from near-$M_\textrm{Ch}$ SNeIa.
The observed Cr/Fe ratio is close to the solar ratio like other cool-core systems \citep[e.g.,][]{Mernier18b}.
When fitted with a model of either near-$M_\textrm{Ch}$ SNIa or sub-$M_\textrm{Ch}$ SNeIa
with a contribution from CCSNe (N20), the $\chi^2$/dof value becomes 14.3/9 and 26.2/9, respectively,
favouring both progenitor types of SNIa substantially contributing to the enrichment.
The contribution of the sub-$M_\textrm{Ch}$ SNeIa is also needed
to explain the observed solar Cr/Fe and Mn/Fe pattern.

Although we use the latest atomic code revised after the {\it Hitomi} observations of the Perseus core,
the observed Ni/Fe ratios of the Centaurus cluster, 1.5--2 times the solar ratio, are 
significantly higher than the solar ratio of the Perseus cluster core
\citep{Simionescu19} and the 44 cool-core systems \citep{Mernier18b}. 
In our Galaxy, both CCSN and SNIa have synthesised Ni in the same way as Fe, 
since the Ni/Fe ratio of stars is $\sim$\,1, with no dependence on
the Fe abundance \citep[][]{FG98,Gratton03}. 
The observed high Ni/Fe ratio suggests that Ni synthesis in the Centaurus cluster core
might be different from that in our Galaxy. However, with CCD detectors,
both Ni-L and K lines are blended into the Fe-L lines and He-like Fe-K$\beta$ line at 7.9\,keV, respectively,
and the derived Ni abundances might be affected by some systematic uncertainties.

\subsubsection{Origin of Nitrogen}
\label{subsubsec:nitrogen}

\begin{figure}
\includegraphics[width=\columnwidth]{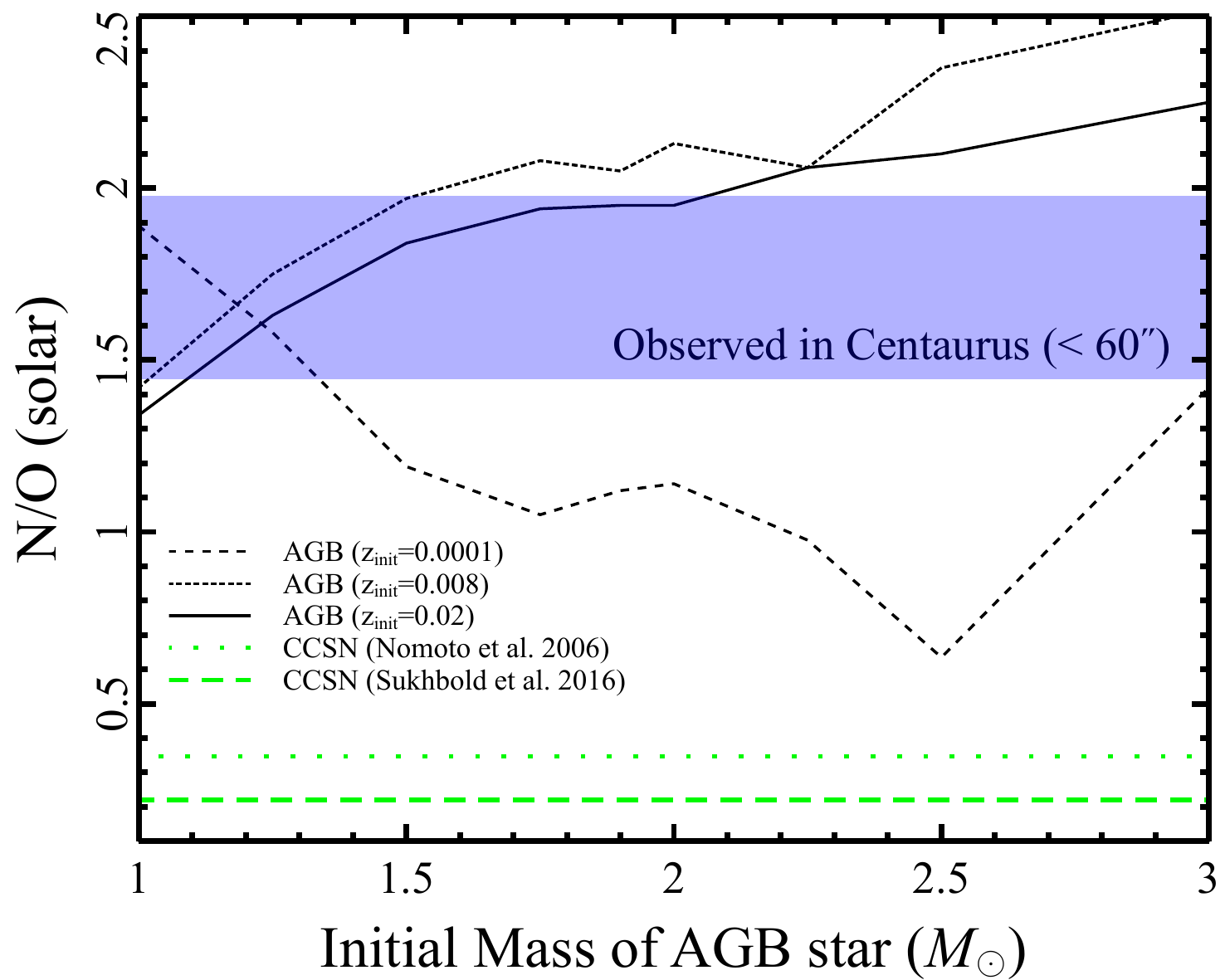}
\caption{The N/O ratio from AGB stars as a function of
initial mass compiled from \citet{Karakas10},
assuming models with several initial metallicity.
The shaded area indicates the observed N/O value with RGS
in the central 60\,arcsec core of the Centaurus cluster.
The CCSN yields from \citet{Nomoto06} and \citet{Sukhbold16}
are also plotted for comparison.
\label{fig:nitrogen}}
\end{figure}

We also detected the N emission line with RGS within $R_\textrm{e}$ of NGC~4696. 
The weighted average of the observed N/O ratio is $\sim$\,1.5--2\,solar
in the central 60\,arcsec region.
As shown in Fig.\,\ref{fig:abund_fit}(a),
the CCSN nucleosynthesis models predict that the N/O ratio is much less than the solar value.
Unlike other light $\alpha$-elements like O and Mg,
the N abundance in the cool cores of some galaxy clusters
and groups, or in early-type galaxies shows super-solar values,
implying a strong chemical contamination by stellar mass loss
from asymptotic giant branch (AGB) stars exhibiting in these systems
\citep[e.g.,][]{Sanders08,Mao19,Mernier22}.
In Fig.\,\ref{fig:nitrogen}, we compare the observed N/O and those expected
from AGB yields by \citet{Karakas10}.
The observed value is  consistent with the calculated ratio assuming AGB stars with
low initial mass (1--2\,solar) and initial metallicity of solar composition.
Thus, the observed N/O ratio indicates the importance of mass loss from low mass AGB stars in the BCG.

\subsection{Comparison with the Stellar Metallicity Profiles}
\label{subsec:metallicity}

In a cool core, mass loss from AGB stars and SNeIa from the BCG continuously supply
metals into  the ICM (see Section\,\ref{sec:intro}).
The former supply metals trapped in stars, which were 
formed from the ISM enriched by early SNe,
and the latter, exploding white dwarfs, pollute the ICM by Fe-peak elements.

\begin{figure}
\includegraphics[width=\columnwidth]{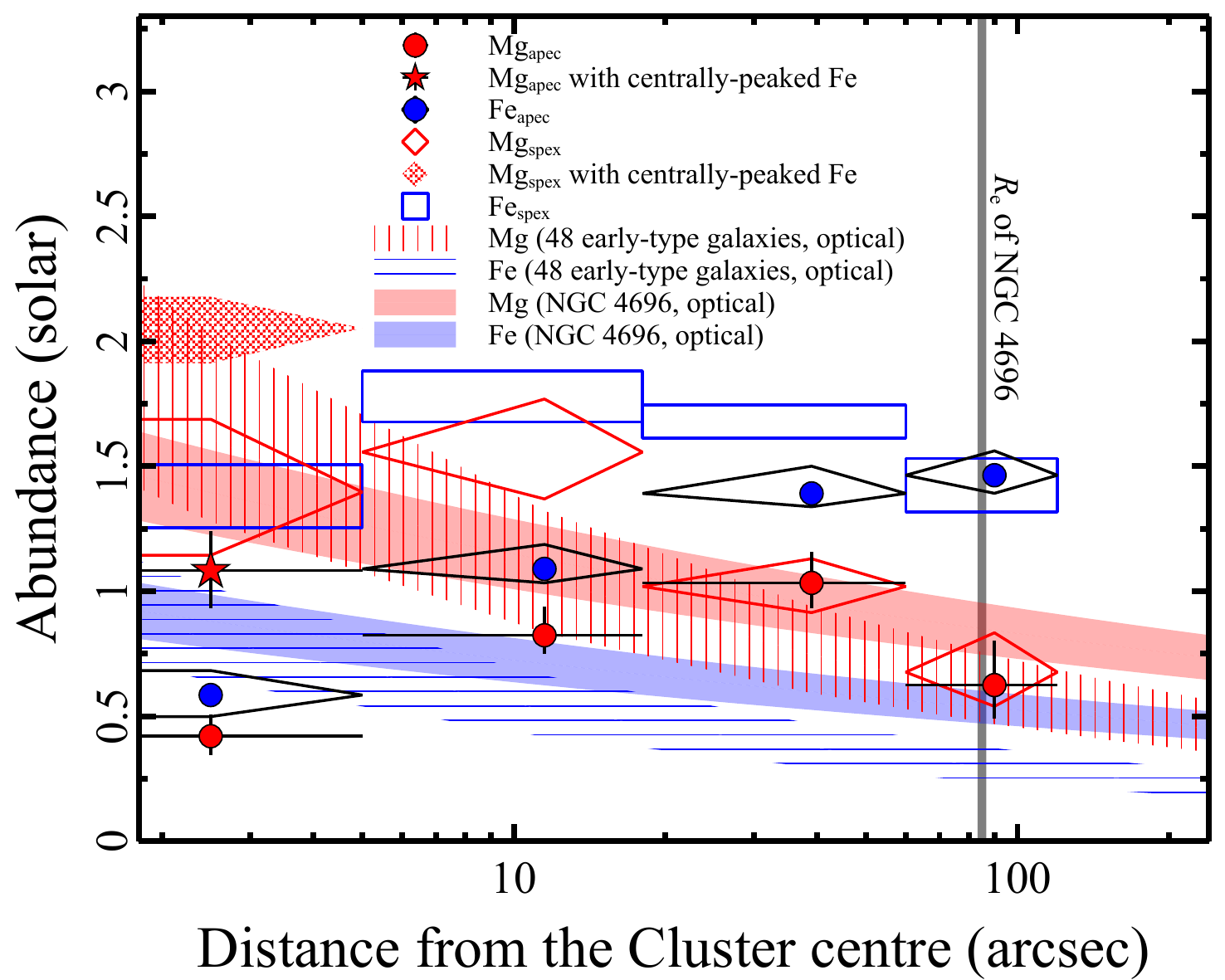}
\caption{The abundance profile of Mg (red) and Fe (blue) overlaid with the stellar metallicity
observed in early-type galaxies (hatched strips, \citealt{Kuntschner10}) and NGC~4696
(shaded strips, \citealt{Kobayashi99}).
\label{fig:mg_metal}}
\end{figure}

In Fig.\,\ref{fig:mg_metal}, we plot the stellar metallicity profiles of NGC~4696 by \citet{Kobayashi99}
and the average profiles of stars in early-type galaxies scaled with $R_\textrm{e}$
by \citet{Kuntschner10}. Here, we adopt $R_\textrm{e} \sim$\,85\,arcsec of NGC~4696
\citep{Carollo93}, and the metallicities are converted
using the proto-solar abundance of \citet{Lodders09}.
The stellar metallicity of galaxies generally increases towards the centre \citep{Carollo93}.
At the effective radii, the stellar metallicity of $\alpha$-elements of these galaxies
are typically close to the solar value,
although  the accurate metallicity  could be different 
by a factor of two because of high uncertainty and 
poor understanding in converts to the metalicity of Mg or Fe
from strength and depth of optical absorption lines \citep{Kobayashi99}.

In Fig.\,\ref{fig:mg_metal}, we also plot 
the Mg and Fe profiles in the ICM of the Centaurus cluster derived from the RGS observations.
As described in the previous subsection, at $R_\textrm{e}$ of the BCG,
the accurate abundances are more reliable than the innermost regions.
We found that the Mg abundances derived from X-ray
measurements around $R_\textrm{e}$ are consistent with the stellar Mg metallicity.
At the inner regions, if we adopt the high Fe abundance model without the abundance drop
described in \S\,\ref{subsec:drop_origin}, i.e. the model with Fe $\sim$\,1.7\,solar,
the  Mg abundances with AtomDB and SPEXACT increase to $\sim$\,1\,solar,
and becomes consistent with the stellar metallicity as shown in Fig.\,\ref{fig:mg_metal}.

When integrating the electron density derived from the VEM profiles out to 120\,arcsec,
the hot gas mass becomes 7$\times 10^{10}\,\textrm{M}_\odot$.
This mass is orders of magnitude higher than those in early-type galaxies.
For example, that of a massive elliptical galaxy
NGC~1404 is only $\sim 7\times 10^{8}\,\textrm{M}_\odot$ \citep{Mernier22}.
Considering the $K$-band luminosity of NGC~4696 is
1.8$\times 10^{12}\,\textrm{L}_\odot$\footnote{Retrieved from Hyperleda: \url{http://atlas.obs-hp.fr/hyperleda/}},
and stellar mass-to-light ratio with $K$-band of giant early-type galaxies are around unity
\citep[e.g.,][]{NM09}, this gas mass is about 4 per cent of the stellar mass of the BCG.
Given that stars lose 10--20 per cent of the initial mass via stellar mass loss
over the Hubble time \citep[e.g.][]{Courteau14,Hopkins22},
the accumulating time of the mass loss of NGC~4696 may be much longer than those of early-type galaxies
while early mass-loss products would have been lost.
Adopting the injection rate of stellar winds from \citet{Ciotti91},
which is $\sim 1.3\,(t/t_\textrm{now})^{-1.3}\,\textrm{M}_\odot$\,yr$^{-1}$ in NGC~4696,
and assuming star formation epoch, for instance, at 2.5\,Gyr ($z\sim$\,2.5,
the peak star formation is at $z=$\,2--3, e.g., \citealt{Madau96,BH05}), 
the time to congregate the observed ICM mass
in the central 120\,arcsec region is about 9--10\,Gyr.
We found no significant change in this estimated time
when adopting different star formation epochs $z=2$ and $z=3$.
In contrast, such a long accumulating time could strongly overestimate the gas mass
in normal early-type galaxies \citep[e.g.,][]{Mernier22}.
Since we compare abundances in stars and ICM which could have very different histories from each other,
the actual accumulation time may change from our simple or simplistic estimation.
For the other clusters with more luminous cool cores, with higher ICM masses
and lower metal abundances, it may be difficult to explain the observed
$\alpha$-elements peak with stellar mass loss \citep[e.g.][]{dePlaa06, Simionescu09a, Mernier17}.

\subsection{Fe mass and SNIa rate}
\label{subsec:femass}

While the total gas mass in the Centaurus core can be accounted for by
the mass ejection through stellar winds (see \S\,\ref{subsec:metallicity}),
how about the masses of each element, especially for Fe?
As shown in Fig.\,\ref{fig:mg_metal}, the Fe abundances in the ICM are significantly higher
than the stellar Fe metallicity, especially around $R_\textrm{e}$ of the BCG,
where the absolute abundances are relatively reliable.
In addition, the Mg/Fe ratios in the ICM are also different from the super-solar $\alpha$/Fe ratios
of stars in giant early-type galaxies \citep[e.g.,][]{Kobayashi99, Kuntschner10}.
These results strongly suggest the additional enrichment of Fe by SNeIa.
Then, the Fe abundances in the ISM can constrain the past metal supply by SNeIa.
Adopting the stellar Fe metallicity of $\sim$\,0.5\,solar,
the Fe abundance synthesised by SNeIa becomes $\sim$\,1\,solar.
From the derived electron density and Fe abundance profiles, the total Fe mass from SNeIa
in the ICM out to 120\,arcsec
becomes $(1.1\pm 0.2) \times 10^8$\,$\textrm{M}_\odot$.
We note that the Fe mass ejected from the stellar population
with the metallicity in Fig.\,\ref{fig:mg_metal} is about $6\times 10^{7}\,\textrm{M}_\odot$
which is a factor of two lower than the observed Fe mass.

The present metal supply by SNeIa can be studied using the Fe abundance in the ISM in early-type galaxies.
The mass of the hot ISM in giant early-type galaxies are typically less than 1 per cent of stellar mass
and the timescale for accumulation of the hot ISM is smaller than $\sim$\,1\,Gyr \citep[][]{Matsushita01}. 
At a given epoch, the absolute abundance of Fe synthesised by SNeIa is calculated by\\
\begin{eqnarray}
M^\textrm{Fe}_\textrm{SNIa} \frac{\theta_\textrm{SNIa}}{\alpha_\ast z^\textrm{Fe}_\textrm{solar}}
\label{eq:1}
\end{eqnarray}\\
Detailed discussion for equation\,\eqref{eq:1} is given in \citet{Matsushita03}.
Here, $M^\textrm{Fe}_\textrm{SNIa}$, the Fe mass forged by one SNIa,
is $\sim 0.5$--$0.7\,\textrm{M}_\odot$ \citep[e.g.,][]{Iwamoto99,Seitenzahl13,Shen18,LN18}, 
$\theta_\textrm{SNIa}$ is the SNIa rate,
$\alpha_\ast$ is the stellar mass-loss rate, and $z^\textrm{Fe}_\textrm{solar} \sim$\,0.001
is the solar Fe mass fraction.
The optical observed rate of SNeIa is 0.1--0.4\,SNIa\,(100\,yr\,$10^{10}\,\textrm{L}_{B,\odot}$)$^{-1}$
\citep[e.g.,][]{Li11}.
Using the injection rate of stellar wind by \citet{Ciotti91}, the expected Fe abundance from SNeIa
in the hot ISM in early-type galaxies is at least 2\,solar \citep[e.g.,][]{Matsushita00,Konami14}.
However, the observed Fe abundances in the ISM of early-type galaxies
with {\it Suzaku} are about 1\,solar \citep[e.g.,][]{Konami14}.
Since the temperatures of the hot ISM in early-type galaxies are close to that of the ICM
in the innermost region of the Centaurus cluster,
there may be some systematic uncertainties in the derived metal abundances in the ISM.

The Fe abundance in the ICM at the Fe-peak of the Centaurus cluster
is significantly higher than those in the hot ISM in early-type galaxies.
We integrated the stellar mass-loss rate and SNIa rate over the past 9--10\,Gyr,
using the mass-loss rate by \citet{Ciotti91} and the delay-time distribution of  SNIa rate modelled
by a power-law behaviour ($\propto t^{-1.34}$, \citealt{Heringer19}).
Assuming star formation at 2.5\,Gyr ($z\sim$\,2.5),
the present SNIa rate is estimated to be 0.11\,SNIa\,(100\,yr\,$10^{10}\,\textrm{L}_{B,\odot}$)$^{-1}$
in order to produce the entire Fe mass observed in the core over the past 9--10\,Gyr.
This value is marginally consistent with the SNeIa rate by the optical observations \citep{Li11}.
When applying star formation at $z=2$ and $z=3$, we did not find remarkable changes in these results.

The ratio of the ICM mass to the stellar mass in the core of the Centaurus cluster
($\sim$\,4 per cent) is a few orders of magnitude higher than those in early-type galaxies,
indicating a longer enrichment time scale (see \S\,\ref{subsec:metallicity}).
In contrast, the iron-mass-to-light ratio of the Centaurus cluster core
($\sim 10^{-5}\,\textrm{M}_\odot/\textrm{L}_{K,\odot}$) is much lower
than those of clusters of galaxies ($\sim 10^{-3}\,\textrm{M}_\odot/\textrm{L}_{K,\odot}$,
\citealt{Matsushita13a,Matsushita13b}).
Our discussed enrichments by SNeIa and/or stellar winds would suggest
a relatively recent or even an ongoing enrichment scenario in the Centaurus core.
However, such a recent enrichment scenario hardly explains the uniform and solar X/Fe ratio
(in early-type galaxies, the Centaurus core, and other clusters)
unless SNeIa and stellar winds have been sophisticatedly controlled
to reproduce the observed abundance pattern, which might be a \textit{handiwork}.
Recently, \citet{Mernier22} discussed that the abundance patterns
observed in cluster or galaxy centres are more likely produced
by an early enrichment which would have started at $z\sim 3$ and ended around $z=2$.
If this is the case, enrichment channels must be CCSNe, the earliest SNeIa, and stellar winds
whose rates are more difficult to be estimated accurately than those at the present.
We also note that such an early enrichment scenario would be challenged
by the escaping or depleting metals that must have been produced subsequently
unless the recent SNeIa and stellar winds are completely quiescent.
Further studies of metal abundance, distribution, and cycle process, not only at central cores
but also at large scales will provide important clues to solving this cosmic conundrum.

\subsection{Future prospects}

Future X-ray mission with high-resolution and spatially resolved spectroscopy
like XRISM is more crucial to reduce the systematic uncertainties
and to estimate the absolute abundance in the cool innermost region of clusters.
Since the observational data, the nucleosynthesis theories, and the atomic codes
are limited at present, we comprehend, as also discussed in \S\,\ref{subsec:drop_origin},
the necessity of extremely high-resolution spectra with micro-calorimeter detectors
onboard future missions like XRISM \citep[][]{XRISM20}
or {\it Athena} \citep[e.g.,][]{Cucchetti18,Mernier20b}.
For example, the Resolve instrument, which will be mounted on XRISM and deliver a constant
$< 7$\,eV spectral resolution, enables us to fully resolve the Ne K-shell emission line from the Fe-L complex.
Offering its non-dispersive X-ray spectroscopy, Resolve provides more accurate spatial
information of emission lines in the cluster and group centres than those with the RGS data. 
By using the low-temperature region of these targets with laboratory experiments
(e.g., electron beam ion traps, \citealt{Gu20}), we expect that the remaining uncertainties
or disagreements between AtomDB and SPEXACT will continue to be reduced.
In particular, our proposed disagreement in the $0.9$--$1.2$\,keV band,
possibly attributed to mismodelling of the \ion{Fe}{XXIII} and \ion{Ne}{X} emission lines
(see \S\,\ref{subsec:drop_origin}), will be an urgent issue to be resolved.
Based on these upcoming leaps, XRISM and/or {\it Athena} will provide crucial knowledge of
metal content in the cluster or group centres,
including the abundance drop, as discussed in \citet{Gastaldello21}.

Furthermore, it is important that the trace odd-$Z$ elements other than N
could also be a sensitive diagnostic of the initial metallicity
of CCSN progenitors \citep[e.g.,][]{Nomoto06,Nomoto13} or
of the stellar population \citep[e.g.,][]{Campbell08,Karakas10}.
Unfortunately, we are not able to constrain the abundance
of these elements while using present grating data.
The micro-calorimeter instruments mounted on {\it Hitomi}
have been shown enough resolution and sensitivity
to constrain the abundance of trace odd-$Z$ elements \citep[e.g.,][]{Simionescu19},
which makes us expect invaluable information about the underlying 
stellar population contributing to the early enrichment history of the ICM.

\section{CONCLUSIONS}
\label{sec:conclusions}

The elemental abundances in the cluster cool cores provide
a key to understanding the chemical enrichment of the central
BCG and the surrounding ICM.
We report the detailed distributions of elements in the ICM expelled by SNe
using the deep {\it Chandra} and {\it XMM-Newton} observations
of the cool core of the Centaurus cluster of galaxies.
While the abundance drop in the innermost region is observed as previous studies
\citep[e.g.,][]{Panagoulia13,Lakhchaura19,Liu19b},
we find that the relative abundance ratios to Fe yield flat profiles towards
the centre, even for noble gas elements, Ne/Fe and Ar/Fe.
In addition, a much higher Fe abundance model without an abundance drop
also reproduces the spectra of the innermost region, excluding the Fe-L complex
around 1.2\,keV with ATOMDB.
These results suggest that the abundance drop may be at least partly artificially
caused by some systematic uncertainties in the atomic data and/or response matrices,
rather than the metal depletion process on to the cold dust grains.

The measured abundance pattern shows that the Si/Fe, S/Fe, Ar/Fe, Ca/Fe, Cr/Fe, and Mn/Fe ratios
are close to the solar values. On the other hand, the O/Fe, Ne/Fe, and Mg/Fe ratios tend to be lower
($\sim$\,0.6--0.9\,solar) and the N/Fe ratio is higher ($\sim$\,1.5\,solar).
Even using the latest versions of atomic code,
the Ni/Fe ratio ($\sim$\,2\,solar) is significantly higher
than the solar ratio and that of the Perseus core \citep[][]{Simionescu19}.
Taking the systematic uncertainties (e.g., atomic codes) into account,
we test linear combination models of recent SN yield calculations for
near- and sub-$M_\textrm{Ch}$ SNIa (\citealt{LN18} and \citealt{Shen18}, respectively),
and CCSN (N20, \citealt{Sukhbold16}; `classical', \citealt{Nomoto06})
in order to reproduce these abundance ratios.
We find that the contribution of both near- and sub-$M_\textrm{Ch}$ SNIa is needed
to explain the Cr/Fe and Mn/Fe (and possibly Ni/Fe) ratios.
The number ratio of CCSNe to the total SNe contributing to the enrichment
is 80 per cent with the latest N20 model and 90 per cent with the `classical' calculation,
consistent with the estimated values in other clusters \citep[e.g.,][]{Simionescu15}.
None the less, the low Ne/Fe and Mg/Fe ratios and the high Ni/Fe ratio are not
reproduced satisfactorily by our linear combination models of each SN population.

Since the N enrichment through SNIa and CCSN is measly,
we compare the observed N/O abundance ratio ($\sim$\,1.7\,solar with RGS)
in the cool core with the calculated values in stellar winds from AGB stars.
We find that the observed N/O ratio is consistent with the values expected in winds
from AGBs with a low initial mass ($\sim 1\textrm{M}_\odot$).
The derived Mg abundance around NGC~4696
is comparable to the stellar metallicity observed in early-type galaxies with optical observations
\citep[][]{Kobayashi99,Kuntschner10}.
These results for the abundance of lighter elements imply
the gas expelled by stellar mass loss is dominant around NGC~4696.
The ICM mass within the central 120\,arcsec core can be reproduced
over the last 9--10\,Gyr enrichment time of stellar mass loss, assuming a star formation at $z=$\,2--3. 
Adopting the power-law delay-time distribution of SNIa rate and accumulating
it over the past 9--10\,Gyr, 
the present SNIa rate required to reproduce the central Fe mass
is 0.11\,SNIa\,(100\,yr\,$10^{10}\,\textrm{L}_{B,\odot}$)$^{-1}$.
This value is marginally consistent with the rate reported by optical observations \citep[e.g.,][]{Li11}.

This study also shows that the high-resolution spectroscopy is important
to measure the metal abundances and to reveal the chemical enrichment
in the cooler innermost region of clusters.
Future missions are crucial to obtain a more precise picture
of the chemical enrichment and evolutionary history of clusters of galaxies
with robust constraints on element abundances especially of Fe-peak and trace odd-$Z$ elements.

\section*{Acknowledgements}

%We sincerely thank the anonymous referee for
%the constructive comments and suggestions,
%which have dramatically improved our manuscript.
We acknowledge financial contribution of Grants-in-Aid for
Scientific Research (KAKENHI) of the Japanese Society
for the Promotion of Science (JSPS) grant No.~16K05300 (KM)
and Grant-in-Aid for JSPS Fellows grant No.~21J21541 (KF).

%%%%%%%%%%%%%%%%%%%%%%%%%%%%%%%%%%%%%%%%%%%%%%%%%%
\section*{Data Availability}

The {\it Chandra} data used in this work are
publicly available from the {\it Chandra} X-ray Center (\url{https://cda.harvard.edu/chaser/}).
The {\it XMM-Newton} Science Archive (\url{http://nxsa.esac.esa.int/nxsa-web/})
stores the observational data used in this paper,
which can be downloaded through the High Energy Astrophysics Science Archive Research Center (HEASARC).
The software packages {\sc heasoft} and {\sc xspec} were used,
and these can be downloaded from the HEASARC software web page
(\url{https://heasarc.gsfc.nasa.gov/docs/software/}).
The figures in this paper were created using {\sc veusz} (\url{https://veusz.github.io/}) and {\sc python} version 3.7.

%The inclusion of a Data Availability Statement is a requirement for articles published in MNRAS. Data Availability Statements provide a standardised format for readers to understand the availability of data underlying the research results described in the article. The statement may refer to original data generated in the course of the study or to third-party data analysed in the article. The statement should describe and provide means of access, where possible, by linking to the data or providing the required accession numbers for the relevant databases or DOIs.



\bibliographystyle{mnras}
\bibliography{centaurus_ref}

\begin{thebibliography}{}
\makeatletter
\relax
\def\mn@urlcharsother{\let\do\@makeother \do\$\do\&\do\#\do\^\do\_\do\%\do\~}
\def\mn@doi{\begingroup\mn@urlcharsother \@ifnextchar [ {\mn@doi@}
  {\mn@doi@[]}}
\def\mn@doi@[#1]#2{\def\@tempa{#1}\ifx\@tempa\@empty \href
  {http://dx.doi.org/#2} {doi:#2}\else \href {http://dx.doi.org/#2} {#1}\fi
  \endgroup}
\def\mn@eprint#1#2{\mn@eprint@#1:#2::\@nil}
\def\mn@eprint@arXiv#1{\href {http://arxiv.org/abs/#1} {{\tt arXiv:#1}}}
\def\mn@eprint@dblp#1{\href {http://dblp.uni-trier.de/rec/bibtex/#1.xml}
  {dblp:#1}}
\def\mn@eprint@#1:#2:#3:#4\@nil{\def\@tempa {#1}\def\@tempb {#2}\def\@tempc
  {#3}\ifx \@tempc \@empty \let \@tempc \@tempb \let \@tempb \@tempa \fi \ifx
  \@tempb \@empty \def\@tempb {arXiv}\fi \@ifundefined
  {mn@eprint@\@tempb}{\@tempb:\@tempc}{\expandafter \expandafter \csname
  mn@eprint@\@tempb\endcsname \expandafter{\@tempc}}}

\bibitem[\protect\citeauthoryear{{Allen} \& {Fabian}}{{Allen} \&
  {Fabian}}{1994}]{AF94}
{Allen} S.~W.,  {Fabian} A.~C.,  1994, \mn@doi [\mnras]
  {10.1093/mnras/269.2.409}, \href
  {https://ui.adsabs.harvard.edu/abs/1994MNRAS.269..409A} {269, 409}

\bibitem[\protect\citeauthoryear{{Arimoto}, {Matsushita}, {Ishimaru}, {Ohashi}
  \& {Renzini}}{{Arimoto} et~al.}{1997}]{Arimoto97}
{Arimoto} N.,  {Matsushita} K.,  {Ishimaru} Y.,  {Ohashi} T.,   {Renzini} A.,
  1997, \mn@doi [\apj] {10.1086/303684}, \href
  {https://ui.adsabs.harvard.edu/abs/1997ApJ...477..128A} {477, 128}

\bibitem[\protect\citeauthoryear{{Arnaud}}{{Arnaud}}{1996}]{Arnaud96}
{Arnaud} K.~A.,  1996, in {Jacoby} G.~H.,  {Barnes} J.,  eds,  Astronomical
  Society of the Pacific Conference Series Vol. 101, Astronomical Data Analysis
  Software and Systems V. p.~17

\bibitem[\protect\citeauthoryear{{Biffi}, {Mernier}  \& {Medvedev}}{{Biffi}
  et~al.}{2018}]{Biffi18}
{Biffi} V.,  {Mernier} F.,   {Medvedev} P.,  2018, \mn@doi [\ssr]
  {10.1007/s11214-018-0557-7}, \href
  {https://ui.adsabs.harvard.edu/abs/2018SSRv..214..123B} {214, 123}

\bibitem[\protect\citeauthoryear{{B{\"o}hringer}, {Matsushita}, {Churazov},
  {Finoguenov}  \& {Ikebe}}{{B{\"o}hringer} et~al.}{2004}]{Boehringer04}
{B{\"o}hringer} H.,  {Matsushita} K.,  {Churazov} E.,  {Finoguenov} A.,
  {Ikebe} Y.,  2004, \mn@doi [\aap] {10.1051/0004-6361:20040047}, \href
  {https://ui.adsabs.harvard.edu/abs/2004A&A...416L..21B} {416, L21}

\bibitem[\protect\citeauthoryear{{Brandt} \& {Hasinger}}{{Brandt} \&
  {Hasinger}}{2005}]{BH05}
{Brandt} W.~N.,  {Hasinger} G.,  2005, \mn@doi [\araa]
  {10.1146/annurev.astro.43.051804.102213}, \href
  {https://ui.adsabs.harvard.edu/abs/2005ARA&A..43..827B} {43, 827}

\bibitem[\protect\citeauthoryear{{Brickhouse}, {Dupree}, {Edgar}, {Liedahl},
  {Drake}, {White}  \& {Singh}}{{Brickhouse} et~al.}{2000}]{Brickhouse00}
{Brickhouse} N.~S.,  {Dupree} A.~K.,  {Edgar} R.~J.,  {Liedahl} D.~A.,  {Drake}
  S.~A.,  {White} N.~E.,   {Singh} K.~P.,  2000, \mn@doi [\apj]
  {10.1086/308350}, \href
  {https://ui.adsabs.harvard.edu/abs/2000ApJ...530..387B} {530, 387}

\bibitem[\protect\citeauthoryear{{Buote} \& {Canizares}}{{Buote} \&
  {Canizares}}{1997}]{Buote97}
{Buote} D.~A.,  {Canizares} C.~R.,  1997, \mn@doi [\apj] {10.1086/303490},
  \href {https://ui.adsabs.harvard.edu/abs/1997ApJ...474..650B} {474, 650}

\bibitem[\protect\citeauthoryear{{Campbell} \& {Lattanzio}}{{Campbell} \&
  {Lattanzio}}{2008}]{Campbell08}
{Campbell} S.~W.,  {Lattanzio} J.~C.,  2008, \mn@doi [\aap]
  {10.1051/0004-6361:200809597}, \href
  {https://ui.adsabs.harvard.edu/abs/2008A&A...490..769C} {490, 769}

\bibitem[\protect\citeauthoryear{{Carollo}, {Danziger}  \& {Buson}}{{Carollo}
  et~al.}{1993}]{Carollo93}
{Carollo} C.~M.,  {Danziger} I.~J.,   {Buson} L.,  1993, \mn@doi [\mnras]
  {10.1093/mnras/265.3.553}, \href
  {https://ui.adsabs.harvard.edu/abs/1993MNRAS.265..553C} {265, 553}

\bibitem[\protect\citeauthoryear{{Cash}}{{Cash}}{1979}]{Cash79}
{Cash} W.,  1979, \mn@doi [\apj] {10.1086/156922}, \href
  {https://ui.adsabs.harvard.edu/abs/1979ApJ...228..939C} {228, 939}

\bibitem[\protect\citeauthoryear{{Chen}, {Wang}, {Zhang}, {Zhang}  \&
  {Ji}}{{Chen} et~al.}{2018}]{Chen18}
{Chen} Y.,  {Wang} Q.~D.,  {Zhang} G.-Y.,  {Zhang} S.,   {Ji} L.,  2018,
  \mn@doi [\apj] {10.3847/1538-4357/aaca32}, \href
  {https://ui.adsabs.harvard.edu/abs/2018ApJ...861..138C} {861, 138}

\bibitem[\protect\citeauthoryear{{Churazov}, {Forman}, {Jones}  \&
  {B{\"o}hringer}}{{Churazov} et~al.}{2003}]{Churazov03}
{Churazov} E.,  {Forman} W.,  {Jones} C.,   {B{\"o}hringer} H.,  2003, \mn@doi
  [\apj] {10.1086/374923}, \href
  {https://ui.adsabs.harvard.edu/abs/2003ApJ...590..225C} {590, 225}

\bibitem[\protect\citeauthoryear{{Ciotti}, {D'Ercole}, {Pellegrini}  \&
  {Renzini}}{{Ciotti} et~al.}{1991}]{Ciotti91}
{Ciotti} L.,  {D'Ercole} A.,  {Pellegrini} S.,   {Renzini} A.,  1991, \mn@doi
  [\apj] {10.1086/170289}, \href
  {https://ui.adsabs.harvard.edu/abs/1991ApJ...376..380C} {376, 380}

\bibitem[\protect\citeauthoryear{{Courteau} et~al.,}{{Courteau}
  et~al.}{2014}]{Courteau14}
{Courteau} S.,  et~al., 2014, \mn@doi [Reviews of Modern Physics]
  {10.1103/RevModPhys.86.47}, \href
  {https://ui.adsabs.harvard.edu/abs/2014RvMP...86...47C} {86, 47}

\bibitem[\protect\citeauthoryear{{Crawford}, {Hatch}, {Fabian}  \&
  {Sanders}}{{Crawford} et~al.}{2005}]{Crawford05}
{Crawford} C.~S.,  {Hatch} N.~A.,  {Fabian} A.~C.,   {Sanders} J.~S.,  2005,
  \mn@doi [\mnras] {10.1111/j.1365-2966.2005.09463.x}, \href
  {https://ui.adsabs.harvard.edu/abs/2005MNRAS.363..216C} {363, 216}

\bibitem[\protect\citeauthoryear{{Cucchetti} et~al.,}{{Cucchetti}
  et~al.}{2018}]{Cucchetti18}
{Cucchetti} E.,  et~al., 2018, \mn@doi [\aap] {10.1051/0004-6361/201833927},
  \href {https://ui.adsabs.harvard.edu/abs/2018A&A...620A.173C} {620, A173}

\bibitem[\protect\citeauthoryear{{De Grandi} \& {Molendi}}{{De Grandi} \&
  {Molendi}}{2001}]{deGrandi01}
{De Grandi} S.,  {Molendi} S.,  2001, \mn@doi [\apj] {10.1086/320098}, \href
  {https://ui.adsabs.harvard.edu/abs/2001ApJ...551..153D} {551, 153}

\bibitem[\protect\citeauthoryear{{Erdim}, {Ezer}, {{\"U}nver}, {Hazar}  \&
  {Hudaverdi}}{{Erdim} et~al.}{2021}]{Erdim21}
{Erdim} M.~K.,  {Ezer} C.,  {{\"U}nver} O.,  {Hazar} F.,   {Hudaverdi} M.,
  2021, \mn@doi [\mnras] {10.1093/mnras/stab2730}, \href
  {https://ui.adsabs.harvard.edu/abs/2021MNRAS.508.3337E} {508, 3337}

\bibitem[\protect\citeauthoryear{{Ettori}}{{Ettori}}{2002}]{Ettori02}
{Ettori} S.,  2002, \mn@doi [\mnras] {10.1046/j.1365-8711.2002.05160.x}, \href
  {https://ui.adsabs.harvard.edu/abs/2002MNRAS.330..971E} {330, 971}

\bibitem[\protect\citeauthoryear{{Feltzing} \& {Gustafsson}}{{Feltzing} \&
  {Gustafsson}}{1998}]{FG98}
{Feltzing} S.,  {Gustafsson} B.,  1998, \mn@doi [\aaps] {10.1051/aas:1998400},
  \href {https://ui.adsabs.harvard.edu/abs/1998A&AS..129..237F} {129, 237}

\bibitem[\protect\citeauthoryear{{Foster}, {Ji}, {Smith}  \&
  {Brickhouse}}{{Foster} et~al.}{2012}]{Foster12}
{Foster} A.~R.,  {Ji} L.,  {Smith} R.~K.,   {Brickhouse} N.~S.,  2012, \mn@doi
  [\apj] {10.1088/0004-637X/756/2/128}, \href
  {https://ui.adsabs.harvard.edu/abs/2012ApJ...756..128F} {756, 128}

\bibitem[\protect\citeauthoryear{{Fruscione} et~al.,}{{Fruscione}
  et~al.}{2006}]{Fruscione06}
{Fruscione} A.,  et~al., 2006, in \procspie. p. 62701V,
  \mn@doi{10.1117/12.671760}

\bibitem[\protect\citeauthoryear{{Fukazawa}, {Ohashi}, {Fabian}, {Canizares},
  {Ikebe}, {Makishima}, {Mushotzky}  \& {Yamashita}}{{Fukazawa}
  et~al.}{1994}]{Fukazawa94}
{Fukazawa} Y.,  {Ohashi} T.,  {Fabian} A.~C.,  {Canizares} C.~R.,  {Ikebe} Y.,
  {Makishima} K.,  {Mushotzky} R.~F.,   {Yamashita} K.,  1994, \pasj, \href
  {https://ui.adsabs.harvard.edu/abs/1994PASJ...46L..55F} {46, L55}

\bibitem[\protect\citeauthoryear{{Gastaldello}, {Simionescu}, {Mernier},
  {Biffi}, {Gaspari}, {Sato}  \& {Matsushita}}{{Gastaldello}
  et~al.}{2021}]{Gastaldello21}
{Gastaldello} F.,  {Simionescu} A.,  {Mernier} F.,  {Biffi} V.,  {Gaspari} M.,
  {Sato} K.,   {Matsushita} K.,  2021, \mn@doi [Universe]
  {10.3390/universe7070208}, \href
  {https://ui.adsabs.harvard.edu/abs/2021Univ....7..208G} {7, 208}

\bibitem[\protect\citeauthoryear{{Gratton}, {Carretta}, {Claudi}, {Lucatello}
  \& {Barbieri}}{{Gratton} et~al.}{2003}]{Gratton03}
{Gratton} R.~G.,  {Carretta} E.,  {Claudi} R.,  {Lucatello} S.,   {Barbieri}
  M.,  2003, \mn@doi [\aap] {10.1051/0004-6361:20030439}, \href
  {https://ui.adsabs.harvard.edu/abs/2003A&A...404..187G} {404, 187}

\bibitem[\protect\citeauthoryear{{Gu} et~al.,}{{Gu} et~al.}{2020}]{Gu20}
{Gu} L.,  et~al., 2020, \mn@doi [\aap] {10.1051/0004-6361/202037948}, \href
  {https://ui.adsabs.harvard.edu/abs/2020A&A...641A..93G} {641, A93}

\bibitem[\protect\citeauthoryear{{Heringer}, {Pritchet}  \& {van
  Kerkwijk}}{{Heringer} et~al.}{2019}]{Heringer19}
{Heringer} E.,  {Pritchet} C.,   {van Kerkwijk} M.~H.,  2019, \mn@doi [\apj]
  {10.3847/1538-4357/ab32dd}, \href
  {https://ui.adsabs.harvard.edu/abs/2019ApJ...882...52H} {882, 52}

\bibitem[\protect\citeauthoryear{{Hitomi Collaboration} et~al.,}{{Hitomi
  Collaboration} et~al.}{2017}]{Hitomi17}
{Hitomi Collaboration} et~al., 2017, \mn@doi [\nat] {10.1038/nature24301},
  \href {https://ui.adsabs.harvard.edu/abs/2017Natur.551..478H} {551, 478}

\bibitem[\protect\citeauthoryear{{Hitomi Collaboration} et~al.,}{{Hitomi
  Collaboration} et~al.}{2018a}]{Hitomi18a}
{Hitomi Collaboration} et~al., 2018a, \mn@doi [\pasj] {10.1093/pasj/psy004},
  \href {https://ui.adsabs.harvard.edu/abs/2018PASJ...70...11H} {70, 11}

\bibitem[\protect\citeauthoryear{{Hitomi Collaboration} et~al.,}{{Hitomi
  Collaboration} et~al.}{2018b}]{Hitomi18b}
{Hitomi Collaboration} et~al., 2018b, \mn@doi [\pasj] {10.1093/pasj/psx156},
  \href {https://ui.adsabs.harvard.edu/abs/2018PASJ...70...12H} {70, 12}

\bibitem[\protect\citeauthoryear{{Hopkins} et~al.,}{{Hopkins}
  et~al.}{2022}]{Hopkins22}
{Hopkins} P.~F.,  et~al., 2022, arXiv e-prints, \href
  {https://ui.adsabs.harvard.edu/abs/2022arXiv220300040H} {p. arXiv:2203.00040}

\bibitem[\protect\citeauthoryear{{Ikebe}, {B{\"o}hringer}  \&
  {Kitayama}}{{Ikebe} et~al.}{2004}]{Ikebe04}
{Ikebe} Y.,  {B{\"o}hringer} H.,   {Kitayama} T.,  2004, \mn@doi [\apj]
  {10.1086/421986}, \href
  {https://ui.adsabs.harvard.edu/abs/2004ApJ...611..175I} {611, 175}

\bibitem[\protect\citeauthoryear{{Iwamoto}, {Brachwitz}, {Nomoto}, {Kishimoto},
  {Umeda}, {Hix}  \& {Thielemann}}{{Iwamoto} et~al.}{1999}]{Iwamoto99}
{Iwamoto} K.,  {Brachwitz} F.,  {Nomoto} K.,  {Kishimoto} N.,  {Umeda} H.,
  {Hix} W.~R.,   {Thielemann} F.-K.,  1999, \mn@doi [\apjs] {10.1086/313278},
  \href {https://ui.adsabs.harvard.edu/abs/1999ApJS..125..439I} {125, 439}

\bibitem[\protect\citeauthoryear{{Johnstone}, {Allen}, {Fabian}  \&
  {Sanders}}{{Johnstone} et~al.}{2002}]{Johnstone02}
{Johnstone} R.~M.,  {Allen} S.~W.,  {Fabian} A.~C.,   {Sanders} J.~S.,  2002,
  \mn@doi [\mnras] {10.1046/j.1365-8711.2002.05743.x}, \href
  {https://ui.adsabs.harvard.edu/abs/2002MNRAS.336..299J} {336, 299}

\bibitem[\protect\citeauthoryear{{Kaastra}}{{Kaastra}}{2017}]{Kaastra17}
{Kaastra} J.~S.,  2017, \mn@doi [\aap] {10.1051/0004-6361/201629319}, \href
  {https://ui.adsabs.harvard.edu/abs/2017A&A...605A..51K} {605, A51}

\bibitem[\protect\citeauthoryear{{Kaastra}, {Mewe}  \&
  {Nieuwenhuijzen}}{{Kaastra} et~al.}{1996}]{Kaastra96}
{Kaastra} J.~S.,  {Mewe} R.,   {Nieuwenhuijzen} H.,  1996, in UV and X-ray
  Spectroscopy of Astrophysical and Laboratory Plasmas. pp 411--414

\bibitem[\protect\citeauthoryear{{Karakas}}{{Karakas}}{2010}]{Karakas10}
{Karakas} A.~I.,  2010, \mn@doi [\mnras] {10.1111/j.1365-2966.2009.16198.x},
  \href {https://ui.adsabs.harvard.edu/abs/2010MNRAS.403.1413K} {403, 1413}

\bibitem[\protect\citeauthoryear{{Katsuda} et~al.,}{{Katsuda}
  et~al.}{2015}]{Katsuda15}
{Katsuda} S.,  et~al., 2015, \mn@doi [\apj] {10.1088/0004-637X/808/1/49}, \href
  {https://ui.adsabs.harvard.edu/abs/2015ApJ...808...49K} {808, 49}

\bibitem[\protect\citeauthoryear{{Kobayashi} \& {Arimoto}}{{Kobayashi} \&
  {Arimoto}}{1999}]{Kobayashi99}
{Kobayashi} C.,  {Arimoto} N.,  1999, \mn@doi [\apj] {10.1086/308092}, \href
  {https://ui.adsabs.harvard.edu/abs/1999ApJ...527..573K} {527, 573}

\bibitem[\protect\citeauthoryear{{Kobayashi} \& {Nomoto}}{{Kobayashi} \&
  {Nomoto}}{2009}]{KN09}
{Kobayashi} C.,  {Nomoto} K.,  2009, \mn@doi [\apj]
  {10.1088/0004-637X/707/2/1466}, \href
  {https://ui.adsabs.harvard.edu/abs/2009ApJ...707.1466K} {707, 1466}

\bibitem[\protect\citeauthoryear{{Konami}, {Matsushita}, {Nagino}  \&
  {Tamagawa}}{{Konami} et~al.}{2014}]{Konami14}
{Konami} S.,  {Matsushita} K.,  {Nagino} R.,   {Tamagawa} T.,  2014, \mn@doi
  [\apj] {10.1088/0004-637X/783/1/8}, \href
  {https://ui.adsabs.harvard.edu/abs/2014ApJ...783....8K} {783, 8}

\bibitem[\protect\citeauthoryear{{Kuntschner} et~al.,}{{Kuntschner}
  et~al.}{2010}]{Kuntschner10}
{Kuntschner} H.,  et~al., 2010, \mn@doi [\mnras]
  {10.1111/j.1365-2966.2010.17161.x}, \href
  {https://ui.adsabs.harvard.edu/abs/2010MNRAS.408...97K} {408, 97}

\bibitem[\protect\citeauthoryear{{Lakhchaura}, {Mernier}  \&
  {Werner}}{{Lakhchaura} et~al.}{2019}]{Lakhchaura19}
{Lakhchaura} K.,  {Mernier} F.,   {Werner} N.,  2019, \mn@doi [\aap]
  {10.1051/0004-6361/201834755}, \href
  {https://ui.adsabs.harvard.edu/abs/2019A&A...623A..17L} {623, A17}

\bibitem[\protect\citeauthoryear{{Leccardi} \& {Molendi}}{{Leccardi} \&
  {Molendi}}{2008}]{LM08}
{Leccardi} A.,  {Molendi} S.,  2008, \mn@doi [\aap]
  {10.1051/0004-6361:200809538}, \href
  {https://ui.adsabs.harvard.edu/abs/2008A&A...486..359L} {486, 359}

\bibitem[\protect\citeauthoryear{{Leung} \& {Nomoto}}{{Leung} \&
  {Nomoto}}{2018}]{LN18}
{Leung} S.-C.,  {Nomoto} K.,  2018, \mn@doi [\apj] {10.3847/1538-4357/aac2df},
  \href {https://ui.adsabs.harvard.edu/abs/2018ApJ...861..143L} {861, 143}

\bibitem[\protect\citeauthoryear{{Li} et~al.,}{{Li} et~al.}{2011}]{Li11}
{Li} W.,  et~al., 2011, \mn@doi [\mnras] {10.1111/j.1365-2966.2011.18160.x},
  \href {https://ui.adsabs.harvard.edu/abs/2011MNRAS.412.1441L} {412, 1441}

\bibitem[\protect\citeauthoryear{{Liu}, {Zhai}  \& {Tozzi}}{{Liu}
  et~al.}{2019}]{Liu19b}
{Liu} A.,  {Zhai} M.,   {Tozzi} P.,  2019, \mn@doi [\mnras]
  {10.1093/mnras/stz533}, \href
  {https://ui.adsabs.harvard.edu/abs/2019MNRAS.485.1651L} {485, 1651}

\bibitem[\protect\citeauthoryear{{Lodders}, {Palme}  \& {Gail}}{{Lodders}
  et~al.}{2009}]{Lodders09}
{Lodders} K.,  {Palme} H.,   {Gail} H.~P.,  2009, \mn@doi [Landolt
  B\"{o}rnstein] {10.1007/978-3-540-88055-4_34}, \href
  {https://ui.adsabs.harvard.edu/abs/2009LanB...4B..712L} {4B, 712}

\bibitem[\protect\citeauthoryear{{Madau}, {Ferguson}, {Dickinson},
  {Giavalisco}, {Steidel}  \& {Fruchter}}{{Madau} et~al.}{1996}]{Madau96}
{Madau} P.,  {Ferguson} H.~C.,  {Dickinson} M.~E.,  {Giavalisco} M.,  {Steidel}
  C.~C.,   {Fruchter} A.,  1996, \mn@doi [\mnras] {10.1093/mnras/283.4.1388},
  \href {https://ui.adsabs.harvard.edu/abs/1996MNRAS.283.1388M} {283, 1388}

\bibitem[\protect\citeauthoryear{{Mao} et~al.,}{{Mao} et~al.}{2019}]{Mao19}
{Mao} J.,  et~al., 2019, \mn@doi [\aap] {10.1051/0004-6361/201730931}, \href
  {https://ui.adsabs.harvard.edu/abs/2019A&A...621A...9M} {621, A9}

\bibitem[\protect\citeauthoryear{{Matsushita}}{{Matsushita}}{2001}]{Matsushita01}
{Matsushita} K.,  2001, \mn@doi [\apj] {10.1086/318389}, \href
  {https://ui.adsabs.harvard.edu/abs/2001ApJ...547..693M} {547, 693}

\bibitem[\protect\citeauthoryear{{Matsushita}}{{Matsushita}}{2011}]{Matsushita11}
{Matsushita} K.,  2011, \mn@doi [\aap] {10.1051/0004-6361/200913432}, \href
  {https://ui.adsabs.harvard.edu/abs/2011A&A...527A.134M} {527, A134}

\bibitem[\protect\citeauthoryear{{Matsushita}, {Makishima}, {Rokutanda},
  {Yamasaki}  \& {Ohashi}}{{Matsushita} et~al.}{1997}]{Matsushita97}
{Matsushita} K.,  {Makishima} K.,  {Rokutanda} E.,  {Yamasaki} N.~Y.,
  {Ohashi} T.,  1997, \mn@doi [\apjl] {10.1086/310939}, \href
  {https://ui.adsabs.harvard.edu/abs/1997ApJ...488L.125M} {488, L125}

\bibitem[\protect\citeauthoryear{{Matsushita}, {Ohashi}  \&
  {Makishima}}{{Matsushita} et~al.}{2000}]{Matsushita00}
{Matsushita} K.,  {Ohashi} T.,   {Makishima} K.,  2000, \mn@doi [\pasj]
  {10.1093/pasj/52.4.685}, \href
  {https://ui.adsabs.harvard.edu/abs/2000PASJ...52..685M} {52, 685}

\bibitem[\protect\citeauthoryear{{Matsushita}, {Finoguenov}  \&
  {B{\"o}hringer}}{{Matsushita} et~al.}{2003}]{Matsushita03}
{Matsushita} K.,  {Finoguenov} A.,   {B{\"o}hringer} H.,  2003, \mn@doi [\aap]
  {10.1051/0004-6361:20021791}, \href
  {https://ui.adsabs.harvard.edu/abs/2003A&A...401..443M} {401, 443}

\bibitem[\protect\citeauthoryear{{Matsushita}, {B{\"o}hringer}, {Takahashi}  \&
  {Ikebe}}{{Matsushita} et~al.}{2007}]{Matsushita07}
{Matsushita} K.,  {B{\"o}hringer} H.,  {Takahashi} I.,   {Ikebe} Y.,  2007,
  \mn@doi [\aap] {10.1051/0004-6361:20041577}, \href
  {https://ui.adsabs.harvard.edu/abs/2007A&A...462..953M} {462, 953}

\bibitem[\protect\citeauthoryear{{Matsushita}, {Sato}, {Sakuma}  \&
  {Sato}}{{Matsushita} et~al.}{2013a}]{Matsushita13a}
{Matsushita} K.,  {Sato} T.,  {Sakuma} E.,   {Sato} K.,  2013a, \mn@doi [\pasj]
  {10.1093/pasj/65.1.10}, \href
  {https://ui.adsabs.harvard.edu/abs/2013PASJ...65...10M} {65, 10}

\bibitem[\protect\citeauthoryear{{Matsushita}, {Sakuma}, {Sasaki}, {Sato}  \&
  {Simionescu}}{{Matsushita} et~al.}{2013b}]{Matsushita13b}
{Matsushita} K.,  {Sakuma} E.,  {Sasaki} T.,  {Sato} K.,   {Simionescu} A.,
  2013b, \mn@doi [\apj] {10.1088/0004-637X/764/2/147}, \href
  {https://ui.adsabs.harvard.edu/abs/2013ApJ...764..147M} {764, 147}

\bibitem[\protect\citeauthoryear{{Mernier}, {de Plaa}, {Lovisari}, {Pinto},
  {Zhang}, {Kaastra}, {Werner}  \& {Simionescu}}{{Mernier}
  et~al.}{2015}]{Mernier15}
{Mernier} F.,  {de Plaa} J.,  {Lovisari} L.,  {Pinto} C.,  {Zhang} Y.~Y.,
  {Kaastra} J.~S.,  {Werner} N.,   {Simionescu} A.,  2015, \mn@doi [\aap]
  {10.1051/0004-6361/201425282}, \href
  {https://ui.adsabs.harvard.edu/abs/2015A&A...575A..37M} {575, A37}

\bibitem[\protect\citeauthoryear{{Mernier} et~al.,}{{Mernier}
  et~al.}{2017}]{Mernier17}
{Mernier} F.,  et~al., 2017, \mn@doi [\aap] {10.1051/0004-6361/201630075},
  \href {https://ui.adsabs.harvard.edu/abs/2017A&A...603A..80M} {603, A80}

\bibitem[\protect\citeauthoryear{{Mernier} et~al.,}{{Mernier}
  et~al.}{2018a}]{Mernier18c}
{Mernier} F.,  et~al., 2018a, \mn@doi [\ssr] {10.1007/s11214-018-0565-7}, \href
  {https://ui.adsabs.harvard.edu/abs/2018SSRv..214..129M} {214, 129}

\bibitem[\protect\citeauthoryear{{Mernier} et~al.,}{{Mernier}
  et~al.}{2018b}]{Mernier18a}
{Mernier} F.,  et~al., 2018b, \mn@doi [\mnras] {10.1093/mnrasl/sly080}, \href
  {https://ui.adsabs.harvard.edu/abs/2018MNRAS.478L.116M} {478, L116}

\bibitem[\protect\citeauthoryear{{Mernier} et~al.,}{{Mernier}
  et~al.}{2018c}]{Mernier18b}
{Mernier} F.,  et~al., 2018c, \mn@doi [\mnras] {10.1093/mnrasl/sly134}, \href
  {https://ui.adsabs.harvard.edu/abs/2018MNRAS.480L..95M} {480, L95}

\bibitem[\protect\citeauthoryear{{Mernier} et~al.,}{{Mernier}
  et~al.}{2020a}]{Mernier20a}
{Mernier} F.,  et~al., 2020a, \mn@doi [Astronomische Nachrichten]
  {10.1002/asna.202023779}, \href
  {https://ui.adsabs.harvard.edu/abs/2020AN....341..203M} {341, 203}

\bibitem[\protect\citeauthoryear{{Mernier} et~al.,}{{Mernier}
  et~al.}{2020b}]{Mernier20b}
{Mernier} F.,  et~al., 2020b, \mn@doi [\aap] {10.1051/0004-6361/202038638},
  \href {https://ui.adsabs.harvard.edu/abs/2020A&A...642A..90M} {642, A90}

\bibitem[\protect\citeauthoryear{{Mernier} et~al.,}{{Mernier}
  et~al.}{2022}]{Mernier22}
{Mernier} F.,  et~al., 2022, \mn@doi [\mnras] {10.1093/mnras/stac253}, \href
  {https://ui.adsabs.harvard.edu/abs/2022MNRAS.511.3159M} {511, 3159}

\bibitem[\protect\citeauthoryear{{Million}, {Werner}, {Simionescu}  \&
  {Allen}}{{Million} et~al.}{2011}]{Million11}
{Million} E.~T.,  {Werner} N.,  {Simionescu} A.,   {Allen} S.~W.,  2011,
  \mn@doi [\mnras] {10.1111/j.1365-2966.2011.19664.x}, \href
  {https://ui.adsabs.harvard.edu/abs/2011MNRAS.418.2744M} {418, 2744}

\bibitem[\protect\citeauthoryear{{Mittal} et~al.,}{{Mittal}
  et~al.}{2011}]{Mittal11}
{Mittal} R.,  et~al., 2011, \mn@doi [\mnras]
  {10.1111/j.1365-2966.2011.19634.x}, \href
  {https://ui.adsabs.harvard.edu/abs/2011MNRAS.418.2386M} {418, 2386}

\bibitem[\protect\citeauthoryear{{Nagino} \& {Matsushita}}{{Nagino} \&
  {Matsushita}}{2009}]{NM09}
{Nagino} R.,  {Matsushita} K.,  2009, \mn@doi [\aap]
  {10.1051/0004-6361/200810978}, \href
  {https://ui.adsabs.harvard.edu/abs/2009A&A...501..157N} {501, 157}

\bibitem[\protect\citeauthoryear{{Nakano}, {Murakami}, {Furuta}, {Enoto},
  {Masuyama}, {Shigeyama}  \& {Makishima}}{{Nakano} et~al.}{2017}]{Nakano17}
{Nakano} T.,  {Murakami} H.,  {Furuta} Y.,  {Enoto} T.,  {Masuyama} M.,
  {Shigeyama} T.,   {Makishima} K.,  2017, \mn@doi [\pasj]
  {10.1093/pasj/psx012}, \href
  {https://ui.adsabs.harvard.edu/abs/2017PASJ...69...40N} {69, 40}

\bibitem[\protect\citeauthoryear{{Nakashima}, {Inoue}, {Yamasaki}, {Sofue},
  {Kataoka}  \& {Sakai}}{{Nakashima} et~al.}{2018}]{Nakashima18}
{Nakashima} S.,  {Inoue} Y.,  {Yamasaki} N.,  {Sofue} Y.,  {Kataoka} J.,
  {Sakai} K.,  2018, \mn@doi [\apj] {10.3847/1538-4357/aacceb}, \href
  {https://ui.adsabs.harvard.edu/abs/2018ApJ...862...34N} {862, 34}

\bibitem[\protect\citeauthoryear{{Nomoto}, {Tominaga}, {Umeda}, {Kobayashi}  \&
  {Maeda}}{{Nomoto} et~al.}{2006}]{Nomoto06}
{Nomoto} K.,  {Tominaga} N.,  {Umeda} H.,  {Kobayashi} C.,   {Maeda} K.,  2006,
  \mn@doi [\nphysa] {10.1016/j.nuclphysa.2006.05.008}, \href
  {https://ui.adsabs.harvard.edu/abs/2006NuPhA.777..424N} {777, 424}

\bibitem[\protect\citeauthoryear{{Nomoto}, {Kobayashi}  \& {Tominaga}}{{Nomoto}
  et~al.}{2013}]{Nomoto13}
{Nomoto} K.,  {Kobayashi} C.,   {Tominaga} N.,  2013, \mn@doi [\araa]
  {10.1146/annurev-astro-082812-140956}, \href
  {https://ui.adsabs.harvard.edu/abs/2013ARA&A..51..457N} {51, 457}

\bibitem[\protect\citeauthoryear{{Ohshiro} et~al.,}{{Ohshiro}
  et~al.}{2021}]{Ohshiro21}
{Ohshiro} Y.,  et~al., 2021, \mn@doi [\apjl] {10.3847/2041-8213/abff5b}, \href
  {https://ui.adsabs.harvard.edu/abs/2021ApJ...913L..34O} {913, L34}

\bibitem[\protect\citeauthoryear{{Panagoulia}, {Fabian}  \&
  {Sanders}}{{Panagoulia} et~al.}{2013}]{Panagoulia13}
{Panagoulia} E.~K.,  {Fabian} A.~C.,   {Sanders} J.~S.,  2013, \mn@doi [\mnras]
  {10.1093/mnras/stt969}, \href
  {https://ui.adsabs.harvard.edu/abs/2013MNRAS.433.3290P} {433, 3290}

\bibitem[\protect\citeauthoryear{{Panagoulia}, {Sanders}  \&
  {Fabian}}{{Panagoulia} et~al.}{2015}]{Panagoulia15}
{Panagoulia} E.~K.,  {Sanders} J.~S.,   {Fabian} A.~C.,  2015, \mn@doi [\mnras]
  {10.1093/mnras/stu2469}, \href
  {https://ui.adsabs.harvard.edu/abs/2015MNRAS.447..417P} {447, 417}

\bibitem[\protect\citeauthoryear{{Russell}, {Sanders}  \& {Fabian}}{{Russell}
  et~al.}{2008}]{Russel08}
{Russell} H.~R.,  {Sanders} J.~S.,   {Fabian} A.~C.,  2008, \mn@doi [\mnras]
  {10.1111/j.1365-2966.2008.13823.x}, \href
  {https://ui.adsabs.harvard.edu/abs/2008MNRAS.390.1207R} {390, 1207}

\bibitem[\protect\citeauthoryear{{Sakuma}, {Ota}, {Sato}, {Sato}  \&
  {Matsushita}}{{Sakuma} et~al.}{2011}]{Sakuma11}
{Sakuma} E.,  {Ota} N.,  {Sato} K.,  {Sato} T.,   {Matsushita} K.,  2011,
  \mn@doi [\pasj] {10.1093/pasj/63.sp3.S979}, \href
  {https://ui.adsabs.harvard.edu/abs/2011PASJ...63S.979S} {63, S979}

\bibitem[\protect\citeauthoryear{{Salpeter}}{{Salpeter}}{1955}]{Salpeter55}
{Salpeter} E.~E.,  1955, \mn@doi [\apj] {10.1086/145971}, \href
  {https://ui.adsabs.harvard.edu/abs/1955ApJ...121..161S} {121, 161}

\bibitem[\protect\citeauthoryear{{Sanders} \& {Fabian}}{{Sanders} \&
  {Fabian}}{2002}]{SF02}
{Sanders} J.~S.,  {Fabian} A.~C.,  2002, \mn@doi [\mnras]
  {10.1046/j.1365-8711.2002.05211.x}, \href
  {https://ui.adsabs.harvard.edu/abs/2002MNRAS.331..273S} {331, 273}

\bibitem[\protect\citeauthoryear{{Sanders}, {Fabian}, {Allen}, {Morris},
  {Graham}  \& {Johnstone}}{{Sanders} et~al.}{2008}]{Sanders08}
{Sanders} J.~S.,  {Fabian} A.~C.,  {Allen} S.~W.,  {Morris} R.~G.,  {Graham}
  J.,   {Johnstone} R.~M.,  2008, \mn@doi [\mnras]
  {10.1111/j.1365-2966.2008.12952.x}, \href
  {https://ui.adsabs.harvard.edu/abs/2008MNRAS.385.1186S} {385, 1186}

\bibitem[\protect\citeauthoryear{{Sanders} et~al.,}{{Sanders}
  et~al.}{2016}]{Sanders16}
{Sanders} J.~S.,  et~al., 2016, \mn@doi [\mnras] {10.1093/mnras/stv2972}, \href
  {https://ui.adsabs.harvard.edu/abs/2016MNRAS.457...82S} {457, 82}

\bibitem[\protect\citeauthoryear{{Sato}, {Tokoi}, {Matsushita}, {Ishisaki},
  {Yamasaki}, {Ishida}  \& {Ohashi}}{{Sato} et~al.}{2007}]{Sato07}
{Sato} K.,  {Tokoi} K.,  {Matsushita} K.,  {Ishisaki} Y.,  {Yamasaki} N.~Y.,
  {Ishida} M.,   {Ohashi} T.,  2007, \mn@doi [\apjl] {10.1086/522031}, \href
  {https://ui.adsabs.harvard.edu/abs/2007ApJ...667L..41S} {667, L41}

\bibitem[\protect\citeauthoryear{{Seitenzahl} et~al.,}{{Seitenzahl}
  et~al.}{2013}]{Seitenzahl13}
{Seitenzahl} I.~R.,  et~al., 2013, \mn@doi [\mnras] {10.1093/mnras/sts402},
  \href {https://ui.adsabs.harvard.edu/abs/2013MNRAS.429.1156S} {429, 1156}

\bibitem[\protect\citeauthoryear{{Shen}, {Kasen}, {Miles}  \&
  {Townsley}}{{Shen} et~al.}{2018}]{Shen18}
{Shen} K.~J.,  {Kasen} D.,  {Miles} B.~J.,   {Townsley} D.~M.,  2018, \mn@doi
  [\apj] {10.3847/1538-4357/aaa8de}, \href
  {https://ui.adsabs.harvard.edu/abs/2018ApJ...854...52S} {854, 52}

\bibitem[\protect\citeauthoryear{{Simionescu}, {Werner}, {B{\"o}hringer},
  {Kaastra}, {Finoguenov}, {Br{\"u}ggen}  \& {Nulsen}}{{Simionescu}
  et~al.}{2009}]{Simionescu09a}
{Simionescu} A.,  {Werner} N.,  {B{\"o}hringer} H.,  {Kaastra} J.~S.,
  {Finoguenov} A.,  {Br{\"u}ggen} M.,   {Nulsen} P.~E.~J.,  2009, \mn@doi
  [\aap] {10.1051/0004-6361:200810225}, \href
  {https://ui.adsabs.harvard.edu/abs/2009A&A...493..409S} {493, 409}

\bibitem[\protect\citeauthoryear{{Simionescu}, {Werner}, {Urban}, {Allen},
  {Ichinohe}  \& {Zhuravleva}}{{Simionescu} et~al.}{2015}]{Simionescu15}
{Simionescu} A.,  {Werner} N.,  {Urban} O.,  {Allen} S.~W.,  {Ichinohe} Y.,
  {Zhuravleva} I.,  2015, \mn@doi [\apjl] {10.1088/2041-8205/811/2/L25}, \href
  {https://ui.adsabs.harvard.edu/abs/2015ApJ...811L..25S} {811, L25}

\bibitem[\protect\citeauthoryear{{Simionescu} et~al.,}{{Simionescu}
  et~al.}{2019}]{Simionescu19}
{Simionescu} A.,  et~al., 2019, \mn@doi [\mnras] {10.1093/mnras/sty3220}, \href
  {https://ui.adsabs.harvard.edu/abs/2019MNRAS.483.1701S} {483, 1701}

\bibitem[\protect\citeauthoryear{{Struble} \& {Rood}}{{Struble} \&
  {Rood}}{1999}]{SR99}
{Struble} M.~F.,  {Rood} H.~J.,  1999, \mn@doi [\apjs] {10.1086/313274}, \href
  {https://ui.adsabs.harvard.edu/abs/1999ApJS..125...35S} {125, 35}

\bibitem[\protect\citeauthoryear{{Str{\"u}der} et~al.,}{{Str{\"u}der}
  et~al.}{2001}]{Strueder01}
{Str{\"u}der} L.,  et~al., 2001, \mn@doi [\aap] {10.1051/0004-6361:20000066},
  \href {https://ui.adsabs.harvard.edu/abs/2001A&A...365L..18S} {365, L18}

\bibitem[\protect\citeauthoryear{{Sukhbold}, {Ertl}, {Woosley}, {Brown}  \&
  {Janka}}{{Sukhbold} et~al.}{2016}]{Sukhbold16}
{Sukhbold} T.,  {Ertl} T.,  {Woosley} S.~E.,  {Brown} J.~M.,   {Janka} H.~T.,
  2016, \mn@doi [\apj] {10.3847/0004-637X/821/1/38}, \href
  {https://ui.adsabs.harvard.edu/abs/2016ApJ...821...38S} {821, 38}

\bibitem[\protect\citeauthoryear{{Takahashi} et~al.,}{{Takahashi}
  et~al.}{2009}]{Takahashi09}
{Takahashi} I.,  et~al., 2009, \mn@doi [\apj] {10.1088/0004-637X/701/1/377},
  \href {https://ui.adsabs.harvard.edu/abs/2009ApJ...701..377T} {701, 377}

\bibitem[\protect\citeauthoryear{{Turner} et~al.,}{{Turner}
  et~al.}{2001}]{Turner01}
{Turner} M.~J.~L.,  et~al., 2001, \mn@doi [\aap] {10.1051/0004-6361:20000087},
  \href {https://ui.adsabs.harvard.edu/abs/2001A&A...365L..27T} {365, L27}

\bibitem[\protect\citeauthoryear{{Urban}, {Werner}, {Allen}, {Simionescu}  \&
  {Mantz}}{{Urban} et~al.}{2017}]{Urban17}
{Urban} O.,  {Werner} N.,  {Allen} S.~W.,  {Simionescu} A.,   {Mantz} A.,
  2017, \mn@doi [\mnras] {10.1093/mnras/stx1542}, \href
  {https://ui.adsabs.harvard.edu/abs/2017MNRAS.470.4583U} {470, 4583}

\bibitem[\protect\citeauthoryear{{Verner}, {Ferland}, {Korista}  \&
  {Yakovlev}}{{Verner} et~al.}{1996}]{Verner96}
{Verner} D.~A.,  {Ferland} G.~J.,  {Korista} K.~T.,   {Yakovlev} D.~G.,  1996,
  \mn@doi [\apj] {10.1086/177435}, \href
  {https://ui.adsabs.harvard.edu/abs/1996ApJ...465..487V} {465, 487}

\bibitem[\protect\citeauthoryear{{Weisskopf}, {Brinkman}, {Canizares},
  {Garmire}, {Murray}  \& {Van Speybroeck}}{{Weisskopf}
  et~al.}{2002}]{Weisskopf02}
{Weisskopf} M.~C.,  {Brinkman} B.,  {Canizares} C.,  {Garmire} G.,  {Murray}
  S.,   {Van Speybroeck} L.~P.,  2002, \mn@doi [\pasp] {10.1086/338108}, \href
  {https://ui.adsabs.harvard.edu/abs/2002PASP..114....1W} {114, 1}

\bibitem[\protect\citeauthoryear{{Werner}, {Durret}, {Ohashi}, {Schindler}  \&
  {Wiersma}}{{Werner} et~al.}{2008}]{Werner08}
{Werner} N.,  {Durret} F.,  {Ohashi} T.,  {Schindler} S.,   {Wiersma} R.~P.~C.,
   2008, \mn@doi [\ssr] {10.1007/s11214-008-9320-9}, \href
  {https://ui.adsabs.harvard.edu/abs/2008SSRv..134..337W} {134, 337}

\bibitem[\protect\citeauthoryear{{Werner}, {Urban}, {Simionescu}  \&
  {Allen}}{{Werner} et~al.}{2013}]{Werner13}
{Werner} N.,  {Urban} O.,  {Simionescu} A.,   {Allen} S.~W.,  2013, \mn@doi
  [\nat] {10.1038/nature12646}, \href
  {https://ui.adsabs.harvard.edu/abs/2013Natur.502..656W} {502, 656}

\bibitem[\protect\citeauthoryear{{Willingale}, {Starling}, {Beardmore},
  {Tanvir}  \& {O'Brien}}{{Willingale} et~al.}{2013}]{Willingale13}
{Willingale} R.,  {Starling} R.~L.~C.,  {Beardmore} A.~P.,  {Tanvir} N.~R.,
  {O'Brien} P.~T.,  2013, \mn@doi [\mnras] {10.1093/mnras/stt175}, \href
  {https://ui.adsabs.harvard.edu/abs/2013MNRAS.431..394W} {431, 394}

\bibitem[\protect\citeauthoryear{{XRISM Science Team}}{{XRISM Science
  Team}}{2020}]{XRISM20}
{XRISM Science Team} 2020, arXiv e-prints, \href
  {https://ui.adsabs.harvard.edu/abs/2020arXiv200304962X} {p. arXiv:2003.04962}

\bibitem[\protect\citeauthoryear{{Yamaguchi} et~al.,}{{Yamaguchi}
  et~al.}{2015}]{Yamaguchi15}
{Yamaguchi} H.,  et~al., 2015, \mn@doi [\apjl] {10.1088/2041-8205/801/2/L31},
  \href {https://ui.adsabs.harvard.edu/abs/2015ApJ...801L..31Y} {801, L31}

\bibitem[\protect\citeauthoryear{{Yoshino} et~al.,}{{Yoshino}
  et~al.}{2009}]{Yoshino09}
{Yoshino} T.,  et~al., 2009, \mn@doi [\pasj] {10.1093/pasj/61.4.805}, \href
  {https://ui.adsabs.harvard.edu/abs/2009PASJ...61..805Y} {61, 805}

\bibitem[\protect\citeauthoryear{{Zhang}, {Wang}, {Foster}, {Sun}, {Li}  \&
  {Ji}}{{Zhang} et~al.}{2019}]{Zhang19}
{Zhang} S.,  {Wang} Q.~D.,  {Foster} A.~R.,  {Sun} W.,  {Li} Z.,   {Ji} L.,
  2019, \mn@doi [\apj] {10.3847/1538-4357/ab4a0f}, \href
  {https://ui.adsabs.harvard.edu/abs/2019ApJ...885..157Z} {885, 157}

\bibitem[\protect\citeauthoryear{{de Plaa} et~al.,}{{de Plaa}
  et~al.}{2006}]{dePlaa06}
{de Plaa} J.,  et~al., 2006, \mn@doi [\aap] {10.1051/0004-6361:20053864}, \href
  {https://ui.adsabs.harvard.edu/abs/2006A&A...452..397D} {452, 397}

\bibitem[\protect\citeauthoryear{{de Plaa}, {Werner}, {Bleeker}, {Vink},
  {Kaastra}  \& {M{\'e}ndez}}{{de Plaa} et~al.}{2007}]{dePlaa07}
{de Plaa} J.,  {Werner} N.,  {Bleeker} J.~A.~M.,  {Vink} J.,  {Kaastra} J.~S.,
   {M{\'e}ndez} M.,  2007, \mn@doi [\aap] {10.1051/0004-6361:20066382}, \href
  {https://ui.adsabs.harvard.edu/abs/2007A&A...465..345D} {465, 345}

\bibitem[\protect\citeauthoryear{{den Herder} et~al.,}{{den Herder}
  et~al.}{2001}]{denHerder01}
{den Herder} J.~W.,  et~al., 2001, \mn@doi [\aap] {10.1051/0004-6361:20000058},
  \href {https://ui.adsabs.harvard.edu/abs/2001A&A...365L...7D} {365, L7}

\makeatother
\end{thebibliography}

%\begin{thebibliography}{99}
%\end{thebibliography}

%%%%%%%%%%%%%%%%%%%%%%%%%%%%%%%%%%%%%%%%%%%%%%%%%%

%%%%%%%%%%%%%%%%% APPENDICES %%%%%%%%%%%%%%%%%%%%%

%\appendix



%%%%%%%%%%%%%%%%%%%%%%%%%%%%%%%%%%%%%%%%%%%%%%%%%%

% Don't change these lines
\bsp	% typesetting comment
\label{lastpage}
\end{document}